\documentclass[showpacs,showkeys,nofootinbib,amsfonts,amssymb,floatfix
,twocolumn,tightenlines,superscriptaddress,prc]{revtex4-2}
\pdfoutput=1

\usepackage{graphicx}  
\usepackage{mathtools}
\usepackage{amsmath, amssymb, bm,bbm}   
\usepackage{xcolor}
\usepackage{natbib}
\usepackage{hyperref}

\usepackage{longtable}
\usepackage{csvsimple} 
\usepackage{adjustbox} 

\usepackage{ifthen}

\usepackage{siunitx} 
\usepackage{balance}

\newcounter{biburlnumpenalty}
\newcounter{biburlucpenalty}
\newcounter{biburllcpenalty}

\newcommand*\xbar[1]{%
   \hbox{%
     \vbox{%
       \hrule height 0.5pt 
       \kern0.3ex
       \hbox{%
         \kern-0.1em
         \ensuremath{#1}%
         \kern-0.1em
       }%
     }%
   }%
} 

\def\descouvemont{0.859}
\def\EFTSfactor{$S_{17}(0)=21.0(7)$ eV b}

\def\){\right)} 
\def\({\left(} 
\def\]{\right]} 
\def\[{\left[}

\def\TrAlpha{$^3\mathrm{H}(\alpha,\gamma)^7\mathrm{Li}$}
\def\HeAlpha{$^3\mathrm{He}(\alpha,\gamma)^7\mathrm{Be}$}

\def\pBe{$^7\mathrm{Be}(p,\gamma)^8\mathrm{B}$}
\def\nLi{$^7\mathrm{Li}(n,\gamma)^8\mathrm{Li}$}

\def\pBeHead{$\bm{^7}$B\lowercase{e}$\bm{(p,\gamma)^8}$B}

\def\swaveHead{$\bm{^3S_1}$-$\bm{^3S_1^\star}$}
\def\pwaveHead{$\bm{^3P_2}$-$\bm{^3P_2^\star}$}

\def\EFTstarHead{EFT$_\star$}
\def\EFTgsHead{EFT$_\text{gs}$}

\def\P{\left(\frac{\stackrel{\rightarrow}{\nabla}}{m_c}
-\frac{\stackrel{\leftarrow}{\nabla}}{m_p}\right)}

\begin{document}

\title{Coupled-channels treatment of  \texorpdfstring{\pBeHead}{p-Be7} in effective field theory}

\author{%
Renato Higa}
\email{higa@if.usp.br}
\affiliation{Instituto de F\'isica, Universidade de S\~ao Paulo, R. do Mat\~ao Nr.1371, 05508-090, S\~ao Paulo, SP, Brazil}

\author{%
Pradeepa Premarathna}
\email{psp63@msstate.edu}
\affiliation{Department of Physics \& Astronomy and HPC$^2$ Center for 
Computational Sciences, Mississippi State
University, Mississippi State, MS 39762, USA}

\author{%
Gautam Rupak}
\email{grupak@ccs.msstate.edu}
\affiliation{Department of Physics \& Astronomy and HPC$^2$ Center for 
Computational Sciences, Mississippi State
University, Mississippi State, MS 39762, USA}

\begin{abstract}
The E1 and M1 contributions to $^7\mathrm{Be}(p,\gamma)^8\mathrm{B}$ at low energies are calculated in halo effective field theory. 
The excited $^7\mathrm{Be}^\star$ core is included as an explicit degree of freedom in a coupled-channels calculation.
The E1 transition is calculated up to next-to-next-to-leading order. The leading contribution from M1 transition that 
gives significant contribution in a narrow energy region around the $1^+$ resonance state of $^8$B is included. 
We compare our results with previous halo effective field theory calculations that also included the $^7\mathrm{Be}^\star$ as an 
explicit degree of freedom. We disagree with these previous calculations in both the formal expressions and also in the analysis.  
Bayesian inference of the data gives \EFTSfactor ~when combined with the expected theory error.
 \end{abstract}

\maketitle

\section{Introduction}
\label{sec:Intro}

The Sun is the fuel for life on Earth. 
The fascinating questions on how the Sun behaves and affects our lives 
propelled many scientific endeavours and outcomes like the impressive 
images of NASA's Solar Dynamics Observatory (SDO)~\cite{Pesnell2012} 
released in June of 2020 after a decade of observations~\cite{NASA-SDO}. 
Practically all of the energy released by the Sun comes from nuclear reactions 
taking place at its core, consuming hydrogen through the $pp$-chain and CNO 
cycle~\cite{Adelberger:2010qa,Wiescher:2012aaa,Brune:2015nsv}. 
The rate of these reactions allows the determination of the age, stability, 
chemical composition, and the fate of the Sun along its history line, as 
predicted by the standard solar model 
(SSM)~\cite{PhysRevLett.78.171,Bahcall:2000nu,Bahcall_2005}. 
A few of the reactions have neutrinos as by-product that travel with practically 
no interaction until eventual detection in dedicated observatories. These 
are key reactions for conveying information about the interior of the Sun, 
such as its temperature that agrees with helioseismological measurements 
to a precision better than 0.2\%~\cite{PhysRevLett.78.171}. 
The first solar neutrinos extensively measured and analysed were those from 
the $\beta^{+}$ decay of the ${}^8{\rm B}$ nucleus. Although with a smaller 
flux compared to other solar reactions, ${}^8{\rm B}$ neutrinos carried the 
necessary energy~\cite{Bahcall_2005} to be efficiently detected prior to 2010. 
In recent years, the 
Borexino experiment were able to detect solar neutrinos from electron capture 
of ${}^7{\rm Be}$~\cite{PhysRevLett.101.091302}, proton-proton 
fusion~\cite{Bellini2014} and $pp$-chain~\cite{Agostini2018}, and from 
the CNO cycle~\cite{Agostini:2020mfq}, boosting solar neutrino studies 
to an unprecedented level~\cite{doi:10.1146/annurev-astro-081811-125539}. 

The radiative capture reaction ${}^7{\rm Be}(p,\gamma)^8{\rm B}$ played an 
important role in uncovering the apparent loss of solar neutrino flux. 
The energetic flux of ${}^8{\rm B}$ neutrinos detected by the 
Super-Kamiokande~\cite{Fukuda:2001nj} and Sudbury Neutrino 
Observatory~\cite{Ahmad:2001an} contrasted with the tight predictions from 
the SSM, lending credit to the phenomenon of neutrino oscillations and 
subsequent Nobel Prizes in 2002 and 2015. 
From observations towards precision, questions like matter versus vacuum 
oscillations or mass hierarchy of neutrino flavors can only be achieved if 
the ${}^7{\rm Be}(p,\gamma)^8{\rm B}$ cross section is known around the 
respective Gamow energy $\sim 20$~keV to a precision better than 
3\%~\cite{doi:10.1146/annurev-astro-081811-125539}. 
Despite recent high-quality data on this reaction, and possible 
underestimation of the respective uncertainties~\cite{Cyburt:2004}, 
experiments are difficult and currently limited to energies above 
100~keV. Thus, the required information can only come from theoretical 
low-energy extrapolations of data, which nowadays dominate the respective 
uncertainties~\cite{doi:10.1146/annurev-astro-081811-125539}. 

The ${}^8{\rm B}$ nucleus is a known example of nuclei with an exotic 
structure that defies the well-established nuclear shell model description 
of tight, stable nuclei. It consists of a ${}^7{\rm Be}$ nucleus as a tight 
core, surrounded by a proton barely bound by $\sim 100$~keV. This is one 
order of magnitude smaller than the few MeVs of one-nucleon separation 
energy in a typical nucleus. One, two, and even more of loosely-bound 
nucleons to a tight core and related new phenomena became an important drive 
of nuclear physics in recent years~\cite{Jensen:2004zz}. Most of its 
universal aspects are captured by assuming the core and valence nucleons 
as elementary degrees of freedom interacting via short-range 
interactions~\cite{Jensen:2004zz,Frederico:2012xh,Hammer:2017tjm}. 
Halo/cluster effective field theory (halo EFT for short) relies on this 
dominant cluster structure as starting point, combining few-body techniques 
and quantum field theory to deliver a model-independent and 
systematically-improvable perturbative calculation in terms of a small ratio 
$Q/\Lambda$. The numerator $Q$ sets the momentum scale characteristic of 
shallow binding energies while the denominator $\Lambda$ is a high momentum 
breakdown scale associated with the tightness of the core. 
By construction, halo EFT works better in the lower energy domain, becoming 
a reliable tool for theoretical low-energy extrapolations~\cite{Rupak:2011nk} 
(see~\cite{Hammer:2017tjm} and references therein). 

Halo EFT has been applied to the ${}^7{\rm Be}(p,\gamma)^8{\rm B}$ reaction 
in previous 
studies~\cite{Zhang:2014zsa,Ryberg:2014exa,Zhang:2015ajn,Zhang:2017yqc}. 
Zhang, Nollett, and Phillips~\cite{Zhang:2014zsa} assessed the dominant 
E1 capture at leading order (LO) that includes the proton and the 
$\frac{3}{2}^{-}$ ground and $\frac{1}{2}^{-}$ excited states of the 
${}^7{\rm Be}$ core as degrees of freedom, with the initial interactions 
fixed by the $s$-wave scattering lengths in the total spin $S=2$ and 
$S=1$ channels, and the final bound state 
given in terms of asymptotic normalization coefficients (ANCs) obtained 
from {\em ab initio} variational Monte Carlo calculation. 
Their result for the astrophysical $S$ factor was $\sim 10$\% smaller than the 
recommended value of 
$S_{17}(0)=\SI{20.8+- 1.6}{\eV\barn}$ from ``Solar II"~\cite{Adelberger:2010qa}, 
though within their theoretical 
LO uncertainties. A next-to-leading order (NLO) calculation, with nine 
adjustable parameters, was carried on with a Bayesian analysis in 
subsequent works~\cite{Zhang:2015ajn,Zhang:2017yqc}, with an improved 
agreement [$S_{17}(0)=\SI{21.3+- 0.7}{\eV\barn}$] with the recommended value. Note that the $S_{17}$ value quoted in Refs.~\cite{Zhang:2015ajn,Zhang:2017yqc}  only includes the errors from the fits. The expected  NLO 10\% theory error in that calculation was not included.
Ryberg {\em et al.}~\cite{Ryberg:2014exa} explored the LO association 
between $S_{17}(0)$ and the charge radius of ${}^8{\rm B}$. With different 
values of ANCs as input, their $S_{17}(0)$ results were scattered between 
$\sim 17$--$20\,\si{\eV\barn}$ and found an apparent linear correlation with 
the charge radius. 

In the present work we perform a next-to-next-to-leading order (NNLO) 
analysis of the ${}^7{\rm Be}(p,\gamma)^8{\rm B}$ capture reaction in the 
halo EFT framework. In contrast to previous 
studies~\cite{Zhang:2014zsa,Ryberg:2014exa,Zhang:2015ajn,Zhang:2017yqc}, 
we include the excitation of the ${}^7{\rm Be}$ core in a coupled-channels 
formalism~\cite{Cohen:2004kf,Lensky:2011he}, extending the calculations 
done for the mirror-symmetric case 
${}^7{\rm Li}(n,\gamma)^8{\rm Li}$~\cite{Higa:2020kfs}. 
The same formal discrepancies raised there about previous halo EFT calculations  apply here as well. 
In particular, the discrepancies are in the initial state $s$-wave scattering and in the final 
$p$-wave bound state. We also disagree with the power counting developed in earlier works. 
We estimate the two-body current contribution to E1 transition to be two orders suppressed in the 
perturbative expansion compared to previous estimates~\cite{Zhang:2015ajn,Zhang:2017yqc}, entering at N$^3$LO instead of NLO.  
Besides the E1 capture, we include the leading M1 contributions 
relevant at energies $\sim 600\;{\rm keV}$ around the $1^+$ 
resonance. That allows us to perform EFT fits over a wider range in energy. The fits in Refs.~\cite{Zhang:2015ajn,Zhang:2017yqc} were restricted to energies below the resonance.
Applying Bayesian analysis on the most recent, high-quality capture data we 
obtain $S_{17}(0)\sim \SI{21}{\eV\barn}$, in agreement with the 
recommended value~\cite{Adelberger:2010qa}. 
As in the ${}^7{\rm Li}(n,\gamma)^8{\rm Li}$ case~\cite{Higa:2020kfs}, one also 
identifies an increasingly important $d$-wave contribution at NLO towards 
higher energies while the dynamics of the ${}^7{\rm Be}$ excited state 
are noticeable only at NNLO.

The paper is organized as follows. Sec.~\ref{sec:EFT} defines the strong and 
electromagnetic EFT Lagrangian from where our interactions are obtained. 
We design two versions of EFTs---one simpler, with only the proton and the 
ground state of the ${}^7{\rm Be}$ core as nuclear degrees of freedom, and 
another including the excited state of the core, to be compared with
each other throughout the paper. 
Secs.~\ref{sec:SWaveMix} and \ref{sec:PwaveMix} provide details on how to 
incorporate the excited core in a coupled-channels formulation in $s$- and 
$p$-waves, respectively. It is followed by a 
brief derivation of the ANCs and their relations with the wave function 
renormalization constants in EFT. One closes Sec.~\ref{sec:Theory} presenting 
basic elements of Bayesian analysis necessary for fits of our EFT parameters 
to capture data. 
The pertinent Feynman diagrams and main expressions of our E1 and M1 capture 
reactions are given in Sec.~\ref{sec:Capture}. Preliminary results are shown in 
Sec.~\ref{sec:PowerCounting} with EFT parameters determined from available 
scattering observables and experimentally determined and calculated ANCs, with the purpose 
of setting up the power counting for this reaction. The main results of 
this work are presented in Sec.~\ref{sec:Analysis} where the EFT parameters 
were constrained by the most recent and precise direct capture 
data~\cite{Filippone:1984us,Hammache:2001,Hammache:1998,Baby:2003,Junghans:2010,Strieder:2001} 
using Bayesian inference. One finishes with a summary and concluding remarks 
in Sec.~\ref{sec:Conclusions}. 

\section{Formalism}
\label{sec:Theory}

In this section we develop the formalism for calculating the radiative capture reaction \pBe. The general formalism is very similar to the calculation in the isospin mirror reaction \nLi~\cite{Higa:2020kfs}. The dominant contribution is from a E1 transition from an initial $s$-wave state to a final $p$-wave bound state. There is a M1 transition from a $p$-wave resonance state that is important around the resonance energy. The modification to \nLi ~due to the Coulomb force  needed in this calculation can be obtained from the halo EFT calculations of \HeAlpha, ~\TrAlpha~\cite{Higa:2016igc,Premarathna:2019tup}. The derivation below closely follows that in Ref.~\cite{Higa:2020kfs} with the Coulomb expressions 
from Refs.~\cite{Higa:2016igc,Premarathna:2019tup}. 

\subsection{Interactions}
\label{sec:EFT}

We start the construction of the interactions by identifying the low-energy degrees of freedom. The $^8$B ground state is a shallow state with a binding energy $B=0.1364$ MeV below the proton-$^7$Be threshold~\cite{Huang:2017,Mang:2017}. The spin-parity assignments of the proton and ground state of $^7$Be are $\frac{1}{2}^+$ and $\frac{3}{2}^{-}$, respectively. Thus the $2^+$   ground state of $^8$B is a $p$-wave bound state. The $^8$B ground state can also be represented as a $p$-wave bound state of the proton and the $\frac{1}{2}^-$ excited core of $^7$Be~\cite{Zhang:2013kja,Zhang:2015ajn,Zhang:2017yqc} that we denote as $^7\mathrm{Be}^\star$.    
Capture from initial $s$-wave states through the E1 transition dominates with subleading contributions from initial $d$-wave states that we describe later. Low-energy capture data shows a prominent contribution from the $1^+$ resonance state of $^8$B. This can be described as a M1 transition between initial and final $p$-wave states.

Following the calculation of \nLi~\cite{Higa:2020kfs}, we construct a theory without the $^7\mathrm{Be}^\star$ core as an explicit degree of freedom that we call \EFTgsHead{}. We construct a second theory with  the $^7\mathrm{Be}^\star$ core as an explicit degree of freedom that we call \EFTstarHead{}. We will show later in Sec.~\ref{sec:PowerCounting} that the two EFTs differ in their momentum dependence only at NNLO. First we present the halo EFT without $^7\mathrm{Be}^\star$ contribution, \EFTgsHead{}. The strong interaction Lagrangian is similar to the one for \nLi ~from \cite{Rupak:2011nk,Fernando:2011ts,Higa:2020kfs}:
 \begin{align}\label{eq:EFT}
\mathcal L= {}&N^\dagger\left[i\partial_0+\frac{\nabla^2}{2 m_p}\right] N 
+C^\dagger\left[i\partial_0+\frac{\nabla^2}{2 m_c}\right]C\nonumber \\
{}&+\sum_\zeta{\chi_{[j]}^{(\zeta)}}^\dagger\left[\Delta^{(\zeta)}+h^{(\zeta)}\left(i\partial_0
+\frac{\nabla^2}{2M}\right)\right]\chi_{[j]}^{(\zeta)}\nonumber \\
{}&+\sqrt{\frac{2\pi}{\mu}}\sum_\zeta
\left[
{\chi_{[j]}^{(\zeta)}}^\dagger N^T P_{[j]}^{(\zeta)}C+\operatorname{h.c.}
\right]\, ,
\end{align}
where $N$ represents the $\frac{1}{2}^{+}$ proton with mass $m_p=938.27$ MeV and charge $Z_p=1$, $C$ represents the $\frac{3}{2}^-$ $^7$Be core with mass $m_c=6536.2$ MeV~\cite{Huang:2017,Mang:2017} and charge $Z_c=4$,  $M=m_p+m_c$ is the total mass, and $\mu= m_p m_c/M$ is the reduced mass. We use natural units with $\hbar=1=c$. In the spectroscopic notation $^{2S+1}L_J$, the initial $s$ waves are $^3S_1$ and $^5S_2$. 
The $p$-wave bound state is a combination of $^3P_2$ and $^5P_2$. The $1^+$ $^8$B resonance is a combination of $^3P_1$ and $^5P_1$. The summation in $\zeta$ is over the $s$- and $p$-wave channels: $^3S_1$, $^5S_2$, $^3P_2$, $^5P_2$, $^3P_1$, $^5P_1$~\cite{Rupak:2011nk,Fernando:2011ts,Higa:2020kfs}.
The projectors in the different $\zeta$ channels are given by $P^{(\zeta)}_{[j]}$ in Eq.~(\ref{eq:EFT}), see   Appendix~\ref{sec:Projectors}. There is a sum over the repeated subscript $[j]$ which is a single index or double indices as appropriate.
We gauge the derivatives of the charged particles with minimal substitution to describe both Coulomb interactions and E1 one-body transition operators. In this theory the binding momentum $\gamma=\sqrt{2\mu B}=14.961$ MeV, the inverse Bohr radius $k_C=\alpha_e Z_c Z_p\mu=23.956$ MeV (with $\alpha_e=e^2/(4\pi)=1/137$ the fine structure constant), and momentum $p\lesssim 40$ MeV constitute the low momentum scale $Q\sim\gamma\sim k_C\sim p$. The momentum $p_R=32.15$ MeV associated with the $1^+$ resonance energy 
$E_R=\SI{0.630+-0.003}{\kilo\eV}$~\cite{Junghans:2003bd} is also considered to scale as $Q$. The breakdown momentum scale of the theory can be estimated from the binding momentum of the $^7$Be core $\Lambda\sim 70$ MeV into the constituents $^3$He-$^4$He.

The M1 transition proceeds through the operators
\begin{align}
    g_p\mu_N N^T\left(\frac{\bm\sigma}{2}\cdot\bm B\right)N
   +g_c\mu_NC^T(\bm J\cdot\bm B)C\nonumber\\
   +\left[i\mu_N L_{22}{\chi^{(^5P_2)}_{ij}}^\dagger B_k \chi_l^{(^5P_1)}R_{ijkl}+\operatorname{h.c.}\right]\, ,
   \label{eq:OM1}
\end{align}
where $\bm\sigma$ are the Pauli matrices, $\bm J$ are the angular momentum matrices for spin-3/2 particle,  $\bm B=\bm\nabla\times\bm A$ is the magnetic field, $\mu_N=e/(2m_p)$ the nuclear magneton, $g_p=2\kappa_p$ the proton gyromagnetic ratio, $g_c=2\kappa_c/3$ the $^7$Be gyromagnetic ratio, and $L_{22}$ is a two-body current coupling~\cite{Fernando:2011ts}. We include only the M1 capture contribution from the dominant ${}^5P_1\rightarrow {}^5P_2$ 
channel. We verified that M1 capture in the other channels are subleading. The anomalous magnetic moments~\cite{Stone:2005} are $\kappa_p=\num{2.79284734+-0.00000003}$ and $\kappa_c=\num{-1.398+-0.015}$, respectively. We present our final results for the $S$-factor at threshold to only 3-significant figures. There is an additional M1 contribution from a magnetic photon coupling to the charged particles ``in flight" that we include~\cite{Fernando:2011ts}.    

 The second halo \EFTstarHead{}  with explicit excited $^7\mathrm{Be}^\star$ core can be described with the Lagrangian~\cite{Higa:2020kfs}
 \begin{align}
\label{eq:EFTstar}
\mathcal L_\star = {}&N^\dagger\left[i\partial_0+\frac{\nabla^2}{2 m_p}\right] N 
+C^\dagger\left[i\partial_0+\frac{\nabla^2}{2 m_c}\right]C\nonumber\\
{}&+C_\star^\dagger\left[i\partial_0-E_\star+\frac{\nabla^2}{2 m_c}\right]C_\star \nonumber\\
{}&+\sum_{\zeta,\zeta'}{\chi_{[j]}^{(\zeta)}}^\dagger\left[
\Pi^{(\zeta\zeta')}+t^{(\zeta\zeta')}\left(i\partial_0
+\frac{\nabla^2}{2M}\right)\right]\chi_{[j]}^{(\zeta')}\nonumber\\
{}&+\sqrt{\frac{2\pi}{\mu}}
\sum_{\zeta}\left[
{\chi_{[j]}^{(\zeta)}}^\dagger N^T P_{[j]}^{(\zeta)}C\right.\nonumber\\
{}&+\left.{\chi_{[j]}^{(\zeta)}}^\dagger N^T P_{[j]}^{(\zeta)}C_\star
+\operatorname{h.c.}
\right]\, ,
\end{align}
where the $C_\star$ field represents the excited $^7\mathrm{Be}^\star$ core with excitation energy $E_\star=0.4291$ MeV. The momentum scale $\gamma_\Delta=\sqrt{2\mu E_\star}=26.54$ MeV is assumed to scale as $Q$. The breakdown momentum for this theory would be set by the binding momentum of the $^7\mathrm{Be}^\star$ core as $\Lambda\sim 60$ MeV.   Compared to Eq.~(\ref{eq:EFT}), this theory
has some additional scattering channels $^3S_1^\star$, $^3P_2^\star$. We also include mixing in the $^3S_1$-$^3S_1^\star$ and $^3P_2$-$^3P_2^\star$ channels. This is facilitated by the off-diagonal terms in the inverse free dimer field $\chi^{(\zeta)}_{[j]}$ propagators. We consider mixing only in the spin $S=1$ channels. The excited core does not participate in the $S=2$ channel.
The capture in the spin $S=2$ channel is known to be about 4 times larger than in the $S=1$ channel~\cite{Baye:2000,Tabacaru:2006}. So we treat any possible mixing between these two spin channels to be subleading and do not consider in our calculation. The summation over $\zeta$ is over the appropriate channels, with the projectors $P^{(\zeta)}_{[j]}$ defined in Appendix~\ref{sec:Projectors}.
The M1 contribution in \EFTstarHead{} is given by the operators in Eq.~(\ref{eq:OM1}) and the in-flight captures.

The E1 contribution to  \pBe ~in \EFTgsHead{} and \EFTstarHead{} 
follows a similar calculation as
 the \HeAlpha\ capture  except for some trivial angular momentum algebra. We use the \HeAlpha ~expressions~\cite{Higa:2016igc} with appropriate modifications. The reader should refer to Ref.~\cite{Higa:2016igc} for technical details on evaluating the Feynman diagrams we consider here. The M1 contribution with Coulomb involves a new integral similar to the E1 integrals that are included in Appendix \ref{sec:Integrals}.

\subsection{ \texorpdfstring{\swaveHead}{3S1-3S1*} Coupled Channels }
\label{sec:SWaveMix}

The coupled-channels $s$-wave scattering amplitude is a $2\times2$ matrix. We write the Coulomb-subtracted scattering matrix as 
\begin{align}
    i\mathcal A^{(ab)}(p)=-i\frac{2\pi}{\mu} [C_0(\eta_p)]^2e^{i2\sigma_0} \mathcal D^{(ab)}(E,0)\, ,
\end{align}
where $E=p^2/(2\mu)$ is the center-of-mass (c.m.) energy and the superscripts are the row-column indices of the amplitude matrix. We identify the $^3S_1$ state as channel 1, and the $^3S_1^\star$ state as channel 2. The Coulomb phase shift is $\sigma_l= \operatorname{arg}\Gamma(l+1+i\eta_p)$ with the Sommerfeld parameter 
$\eta_p=k_C/p$. The parameter $[C_0(\eta_p)]^2$, associated with the probability of the Coulomb wave function at the origin, is given by
\begin{align}
    C_l(\eta_p)=\frac{2^l e^{-\pi\eta_p/2}|\Gamma(l+1+i\eta_p)|}{\Gamma(2l+2)}\, .
\end{align}

Following the \nLi~ calculation in Ref.~\cite{Higa:2020kfs}, we write the inverse dimer propagator as 
\begin{align}
    \mathcal D^{-1}= \mathcal D_0^{-1} -\Sigma\, ,
\label{eq:fulldimer}
\end{align}
where $\mathcal D_0^{-1}$ is the inverse free dimer propagator and $\Sigma$ is the self-energy.  Calculation of $[\mathcal D(E,0)]^{-1}$ is simpler than $\mathcal D(E,0)$. We have the free inverse dimer propagator from Eq.~(\ref{eq:EFTstar}):
\begin{align}
    [\mathcal D_0(E,0)]^{-1}&= 
    \begin{bmatrix} \Pi^{(11)}   & 
    \Pi^{(12)} \\
  \Pi^{(12)} & \Pi^{(22)}
    \end{bmatrix}
   \, ,
\end{align}
where we only keep the couplings $\Pi^{(ij)}$ in a low momentum expansion. In a single-channel calculation this would correspond to keeping only the scattering length contribution. The self-energy is 
\begin{align}
   - \Sigma(E,0)&=-\frac{2\pi}{\mu}   
    \begin{bmatrix} J_0(-ip) & 0\\
  0& J_0(-ip_\star)
    \end{bmatrix}
 \, ,
\end{align}
where $p_\star=\sqrt{p^2-\gamma_\Delta^2+i0^+}$ and 
\begin{align}
    J_0(x)={}&-2\mu\int\frac{d^3q}{(2\pi)^3}\frac{1}{q^2+ x^2}\frac{2\pi\eta_q}{e^{2\pi\eta_q}-1}\nonumber \\
    ={}&-\frac{\mu}{2\pi}\lambda +\frac{k_C\mu}{2\pi}\left[\frac{1}{D-4}+1-3\gamma_\mathrm{E}\right.\nonumber\\
&\left.+2\ln\frac{\lambda\sqrt{\pi}}{2k_C}-2H(-ik_C/x) \right]\, .
\end{align}
$H(x)=\psi(ix)+1/(2ix)-\ln(ix)$ with  the di-gamma function $\psi(x)$. 
We regulate the loop integrals using dimensional regularization in the
power divergence subtraction (PDS) scheme~\cite{Kaplan:1998tg} that removes all divergences in space-time dimensions $D\leq 4$. $\lambda$ is the renormalization scale.  
With the RG conditions
\begin{align}
     \Pi^{(ij)}={}\frac{1}{a_{ij}}
     +k_C\left[-\frac{\lambda}{k_C}+\frac{1}{D-4}+1-3\gamma_\mathrm{E}\right.\nonumber\\
     \left.+2\ln\frac{\lambda\sqrt{\pi}}{2k_C} \right]\delta_{ij}\, ,
     \label{eq:RGcond}
\end{align}
where $a_{ij}$s have units of length~\cite{Cohen:2004kf},  we arrive at 
\begin{align}
   \left[ \mathcal D(E,0)\right]^{-1}={}& 
    \begin{bmatrix} 1/a_{11} &
  1/a_{12}\\
  1/a_{12}&
1/a_{22}
    \end{bmatrix}\nonumber\\
    &{}+
    \begin{bmatrix}  2 k_C H\!\left({k_C}/{p}\right)&
  0\\
  0&
2 k_C H\!\left( {k_C}/{p_\star}\right)
    \end{bmatrix}\, ,
\end{align}
which can be used to calculate the $s$-wave scattering matrix $\mathcal A(p)$. More specifically, we calculate
\begin{align}\label{eq:swaveA11}
    \mathcal A^{(11)}={}&\frac{2\pi}{\mu}
    [C_0(\eta_p)]^2e^{i2\sigma_0}
    \Bigg\{-a_{11}^{-1}-2 k_C
   H\left(\frac{k_C}{p}\right) \nonumber\\
   &+a_{12}^{-2}\left[\frac{1}{a_{22}}+2 k_C
   H\left(\frac{ k_C}{p_\star}\right)\right]^{-1}\Bigg\}^{-1}\, ,
\end{align}
which reduces to the single-channel result for $\Pi_{12}=0$. For the off-diagonal contribution,
\begin{multline}\label{eq:swaveA12}
    \mathcal A^{(12)}=\frac{2\pi}{\mu} 
    [C_0(\eta_p)]^2e^{i2\sigma_0}\Bigg\{-a_{12}^{-1}
  +a_{12}\left[\frac{1}{a_{11}}\right.\\
  \left. +2 k_C
   H\left(\frac{k_C}{p}\right)\right]
   \left[\frac{1}{a_{22}}+2 k_C
  H\left(\frac{ k_C}{p_\star}\right)\right]\Bigg\}^{-1}\, .
\end{multline}

Matching $\mathcal A_{11}$ to the effective range expansion (ERE) at $p\ll\gamma_\Delta$ we get
\begin{align}
    -a_{11}^{-1}&-a_{12}^{-2}\left[-a_{22}^{-1}-2 k_C
   H\left(-\frac{i k_C}{\sqrt{-p^2+\gamma
   _{\Delta }^2-i 0^+}}\right)\right]^{-1}\nonumber\\
\approx{}&
     -\frac{1}{a_{11}}+\frac{a_{22}a_{12}^{-2}}{1+2a_{22}k_C H\left(-i\frac{ k_C}{\gamma_\Delta}\right)}\nonumber \\
     &+
i\frac{a_{22}^2 a_{12}^{-2} k_C^2H'\left(-i \frac{k_C}{\gamma_\Delta}\right)}{\gamma_\Delta^3\left[1+2a_{22}H\left(-i\frac{k_C}{\gamma_\Delta}\right)\right]^2}p^2
+\dots\nonumber \\
={}&-\frac{1}{a_0^{(1)}}+\frac{1}{2}r_0^{(1)} p^2+\dots\,, 
\label{eq:EREmatch1}
\end{align}
which gives
\begin{align}\label{eq:EFTstarConstrains}
    a_{11}
    &=a_0^{(1)}\frac{ 1+2 a_{22}
   k_C H\left(-\frac{i k_C}{\gamma
   _{\Delta }}\right)}
   {1+
   a_{22}\left[a_0^{(1)} a_{12}^{-2}+2 k_C
   H\left(-\frac{i k_C}{\gamma
   _{\Delta }}\right)\right]}\, , \nonumber\\
    a_{12}^{-2} 
    &=-r_0^{(1)}\frac{i  \gamma _{\Delta }^3
   \left[1+2 a_{22} k_C
   H\left(-\frac{i k_C}{\gamma_\Delta }\right)\right]^2}{2
   a_{22}^2 k_C^2 H'\left(-\frac{i
   k_C}{\gamma _{\Delta }}\right)}\, .
\end{align}
These expressions reduce to the ones without Coulomb interactions in Ref.~\cite{Higa:2020kfs}. A mixed channel calculation (without the Coulomb force) using nucleon-core contact interaction in $s$-wave, instead of dimer-particle interaction as we do here, was presented in Ref.~\cite{Cohen:2004kf}. See also Ref.~\cite{Lensky:2011he} where a coupled-channels calculation  with the Coulomb force was considered.

In the $S=1$ channel, the new measurement~\cite{Paneru:2019jyz} of  $a_0^{(1)}=17.34^{+1.11}_{-1.33}\, \si{\femto\meter}$ would make the scattering length scale as $1/Q$.  We take all $a_{ij}\sim 1/Q$ which then gives $r_0^{(1)}\sim 1/Q$ for $k_C\sim \SI{24}{\mega\eV}\sim Q$.  Irrespective of the sign of $a_{12}$, $r_0^{(1)}$ is negative, just as it is in the system without Coulomb interactions~\cite{Cohen:2004kf,Higa:2020kfs}. Though we started with a theory with momentum-independent contact interactions, an $s$-wave effective range $r_0^{(1)}$ is generated in the coupled-channels calculation as there is a finite difference in the relative momentum in the scattering between states with and without the excited core~\cite{Higa:2020kfs}.   

The calculations in Eqs.~(\ref{eq:EREmatch1}), (\ref{eq:EFTstarConstrains}) are in fact a matching calculation of EFT$_\text{gs}$ to EFT$_\star$. Eqs. (\ref{eq:RGcond}) and (\ref{eq:EFTstarConstrains}) determine the strong-interaction couplings $\Pi^{(ij)}$ that appear in EFT$_\star$ from the ERE parameters $a^{(1)}_0$, $r^{(1)}_0$, whose relations with the corresponding EFT$_\mathrm{gs}$ couplings are known~\cite{Higa:2020kfs,Griesshammer:2004pe}. More specifically, it shows how the $s$-wave scattering amplitude calculated in EFT$_\star$ is reproduced by EFT$_\text{gs}$ at $p\ll\gamma_\Delta$.

In the analysis we present later, we use Eq.~(\ref{eq:EFTstarConstrains}) to constrain $a_{11}$ by $a_0^{(1)}$. Thus, $a_{22}$ and $a_{12}$ remain the only free parameters to be fitted to determine the $s$-wave scattering amplitudes in the coupled-channels calculation.

\subsection{\texorpdfstring{\pwaveHead}{3P2-3P2*} Coupled Channels}
\label{sec:PwaveMix}
The coupled-channels calculation for the $p$-wave bound state is similar to the $s$-wave scattering 
calculation, 
and follows the derivation in Ref.~\cite{Higa:2020kfs}, with the additional complexity of Coulomb interactions. We write the free inverse dimer propagator as
\begin{align}
    [\mathcal D_0(E,0)]^{-1}&= 
    \begin{bmatrix} \Pi^{(11)} +  E t^{(11)}   & 
    \Pi^{(12)} +  E t^{(12)} \\
  \Pi^{(12)} + E t^{(12)} & \Pi^{(22)} +  E t^{(22)}
    \end{bmatrix}
     \, ,
\end{align}
where we identify $^3P_2$ as channel 1 and $^3P_2^\star$ as channel 2. Here we keep the effective momentum contributions since $p$-wave bound state calculation requires both a momentum-independent and a momentum-dependent interactions at LO~\cite{Bertulani:2002sz,Bedaque:2003wa} in halo EFT.

The $p$-wave self-energy term is given by
\begin{align}
   - \Sigma(E,0)&=-\frac{6\pi}{\mu^3}  
    \begin{bmatrix} J_1(-ip )  & 0\\
  0& J_1(-ip_\star) 
    \end{bmatrix}
    \, ,
\end{align}
where
\begin{align}
    J_1(x)={}&-\frac{\mu k_C}{3\pi}(k_C^2-x^2)H(-i k_C/x)\nonumber\\
    &-\frac{\mu k_c}{3\pi}(k_C^2-x^2) \alpha-\frac{\mu}{3\pi} \beta\, ,\nonumber\\
    \alpha ={}&\frac{1}{D-4}-\frac{4}{3}+\frac{3}{2}\gamma_E-\ln\frac{\lambda\sqrt{\pi}}{2 k_c}+\frac{3\lambda}{4 k_C}\, ,\nonumber\\
    \beta={}& 4\pi^2 k_C^3\zeta'(-2)+\frac{\lambda k_C^2}{4}
    -\frac{3\pi\lambda^2 k_C}{4}+\frac{\pi\lambda^3 }{4}
    \, .
\end{align}
The $p$-wave amplitude is 
\begin{align}
    \mathcal A(p) &=-9 [C_1(\eta_p)]^2 e^{i2\sigma_1}\frac{2\pi}{\mu}\frac{p^2}{\mu^2}\mathcal D(E,0)\, .
\end{align}

We use the RG conditions 
\begin{align}
    \mu^2\Pi^{(11)}={}&a_{11}^{-1}-k_C^3 \left[\frac{3\lambda}{2k_C}+8 \pi ^2 \zeta'(-2)+\frac{2}{D-4}\right.\nonumber\\
    &\left.+3 \gamma
   -\frac{8}{3}-\ln\frac{\pi  \lambda ^2}{4 k_C^2}\right]-\lambda \frac{k_C^2-3 \pi  \lambda  k_C+\pi \lambda ^2}{2}\, ,\nonumber\\
   \mu^2\Pi^{(22)}={}&a_{22}^{-1}- k_C \left(k_C^2-\gamma _{\Delta }^2\right)
   \left[\frac{3 \lambda }{2 k_C}+8 \pi ^2 \zeta'(-2)
  \right.\nonumber\\ &\left.
   +\frac{2}{D-4}+3 \gamma
   -\frac{8}{3}-\ln\frac{\pi  \lambda ^2}{4 k_C^2}\right]
   \nonumber\\
  &  - \lambda  \frac{k_C^2-3 \pi  \lambda  k_C+\pi \lambda ^2}{2}
  -8 \pi ^2 \zeta'(-2)
   k_C\gamma_\Delta^2\nonumber\\
  &-2 k_C \left(k_C^2-\gamma _{\Delta }^2\right)
   H\left(-\frac{i k_C}{\gamma _{\Delta }}\right)\, ,\nonumber\\
    \mu^2 \Pi^{(12)}={}&a_{12}^{-1}\, ,
   \end{align}
   and
   \begin{align}
   \mu t^{(11)}=&-r_{11} -3 \lambda -2 k_C\left[\frac{2}{D-4}+3 \gamma -\frac{8}{3}-
   \ln\frac{\pi  \lambda ^2}{4 k_C^2}\right]\, ,\nonumber\\
   \mu t^{(22)}=&-r_{22} -3 \lambda -2 k_C\left[\frac{2}{D-4}+3 \gamma -\frac{8}{3}-
   \ln\frac{\pi  \lambda ^2}{4 k_C^2}\right]\nonumber\\
   &+\frac{2 i \left(k_C^4-\gamma _{\Delta }^2
   k_C^2\right) H'\left(-\frac{i k_C}{\gamma _{\Delta }}\right)}{\gamma
   _{\Delta }^3}\nonumber\\
   & -4 k_C H\left(-\frac{i k_C}{\gamma _{\Delta }}\right)\, ,\nonumber\\
   \mu t^{(12)}=&-r_{12}\, ,
\end{align}
where $a_{ij}$ are scattering volumes though we used the same notation earlier for scattering lengths in the  $^3S_1$-$^3S_1^\star$ coupled-channels. The scattering volumes do not appear in the cross section formula as we show below, and they are not used outside of this subsection. At $p\ll\gamma_\Delta$, by looking at $\mathcal A^{(11)}(p)$, we get
\begin{align}
    &9[C_1(\eta_p)]^2(p^3\cot\delta_1-ip^3)\nonumber\\
    {}&\approx \frac{a_{22}}{a_{12}^2}-\frac{1}{a_{11}}+
    \frac{p^2 \left(a_{12}^2 r_{11}-2 a_{22} a_{12} r_{12}+a_{22}^2
   r_{22}\right)}{2 a_{12}^2}\nonumber \\
   &{}-2 k_C(k_C^2+p^2) H(\eta_p)\nonumber \\
  {}& =-\frac{1}{a_1^{(1)}}+\frac{1}{2} r_1^{(1)} p^2 -2 k_C(k_C^2+p^2) H(\eta_p)\, .
\label{eq:ERE-pwave}
\end{align}
As for the $s$ wave, this relation is  a matching calculation of EFT$_\text{gs}$ to EFT$_\star$ for the $p$ wave amplitude at low energy.

For the wave function normalization constant we get
\begin{align}
    \mathcal Z^{-1}&=\frac{d}{dE}[\mathcal D(E,0)]^{-1}\Big|_{E=-B}=-\frac{1}{\mu}
\begin{bmatrix}
\rho_{11} & r_{12} \\
r_{12} & \rho_{22} \\
\end{bmatrix}\, ,
\end{align}
with
\begin{align}
\rho_{11}={}&r_{11}\!+\!\frac{2 i k_C^2 \left(k_C^2\!-\!\gamma ^2\right) 
H'\left(-\frac{ik_C}{\gamma }\right)}{\gamma ^3}
\!-\!4 k_C H\left(-\frac{ik_C}{\gamma}\right)\nonumber\\
\approx{}& r_{11}-\SI{1.5538}{\mega\eV}\,,
\nonumber\\
\rho_{22}={}&r_{22}
-\frac{2 i \gamma ^2 k_C^2 H'\left(-\frac{i k_C}{\gamma_\star}\right)}{\gamma_\star^3}
-\frac{2 i k_C^4 H'\left(-\frac{i k_C}{\gamma_{\Delta }}\right)}{\gamma _{\Delta }^3}\nonumber\\
&-\frac{2 i \gamma_{\Delta}^2 k_C^2 H'\left(-\frac{i k_C}{\gamma_\star}\right)}
{\gamma_\star^3}
+\frac{2 i k_C^2 H'\left(-\frac{i k_C}{\gamma_{\Delta}}\right)}{\gamma _{\Delta }}
\nonumber\\
&+\frac{2 i k_C^4 H'\left(-\frac{ik_C}{\gamma_\star}\right)}{\gamma_\star^3}
+4 k_C H\left(-\frac{i k_C}{\gamma _{\Delta }}\right)\nonumber\\
&-4 k_C H\left(-\frac{i k_C}{\gamma_\star}\right)\nonumber\\
\approx{}& r_{22}+\SI{5.0996}{\mega\eV}\,,
\label{eq:Z-ere-relations}
\end{align}
and $\gamma_\star=\sqrt{\gamma^2+\gamma_\Delta^2}$. 
We see from above that $\mathcal Z^{-1}$, and consequently $\mathcal Z$ depends on three effective momenta $r_{11}$, $r_{12}$, $r_{22}$. The two ANCs $C_{1,^3P_2}^2$, $C_{1,^3P_2^\star}^2$, discussed below, are not sufficient to constraint the three $r_{ij}$s. If we assume 
$k_C\sim Q\sim \gamma\sim \gamma_\Delta$, and $r_{ij}\sim\Lambda$, we can write $\mathcal Z$ using only the effective momenta $r_{ij}$s.
In our coupled-channels calculation we simply fit the wave function renormalization constants $\mathcal Z_{11}$, $\mathcal Z_{12}$ to the ANCs without attempting to interpret these in terms of the $r_{ij}$s~\cite{Higa:2020kfs}.

\subsection{Asymptotic Normalization Constant}
\label{sec:ANC}

Here we present a derivation of the relation between the ANCs and the wave 
function renormalization constants. We follow the convention from 
Refs.~\cite{Hamilton:1973xd,vHaeringen77,deMaag:1983kpl,Iwinski:1984rj,Sparenberg:2009rv,Yarmukhamedov:2011kd}. 
Suppressing spin indices, the $S$-matrix projected onto $l$th partial wave is 
\begin{align}
S_{l}(p^2)={}&e^{2i\sigma_{l}}e^{2i\delta_{l}}\nonumber\\
={}&2ip\,F_{l}(p^2)\,\frac{p^{2l}\Gamma(l\!+\!1\!+\!i\eta_p)^2}
{e^{\pi\eta_p}\Gamma(l\!+\!1)^2}+e^{2i\sigma_{l}}\,,
\end{align}
where $\delta_{l}$ is the Coulomb-subtracted partial wave phase shift. 
The function $F_{l}$ relates to the elastic scattering amplitude via 
\begin{align}
{\cal A}_{l}(p^2)={}&\frac{i\pi}{\mu p}\,e^{2i\sigma_{l}}(e^{2i\delta_{l}}-1)\nonumber\\
={}&-\frac{2\pi}{\mu}\frac{p^{2l}\Gamma(l\!+\!1\!+\!i\eta_p)^2e^{-\pi\eta_p}}
{\Gamma(l\!+\!1)^2}F_{l}(p^2)\,.
\end{align}
The generalization of the Landau-Smorodinsky $K$ function, analytic around 
$p\sim 0$, is obtained from 
$F_{l}(p^2)$~\cite{Yarmukhamedov:2011kd,Sparenberg:2009rv,Iwinski:1984rj}, 
\begin{align}
K_{l}(p^2)&=\frac{1}{F_{l}(p^2)}+p^{2l+1}d_{l}(\eta_p)\,2\eta_p H(\eta_p)\nonumber\\
&=p^{2l+1}d_{l}(\eta_p)\left[C^2_0(\eta_p)(\cot\delta_l-i)
+2\eta_p H(\eta_p)\right]
\nonumber\\
&\approx
-\frac{1}{a_{l}}+r_{l}\frac{p^2}{2}+{\cal P}_{l}\frac{p^4}{4}+\cdots \,,
\label{eq:EREdef1}
\end{align}
where the last equality is the Coulomb ERE
with the conventions of 
Refs.~\cite{Hamilton:1973xd,vHaeringen77,deMaag:1983kpl,Iwinski:1984rj,Sparenberg:2009rv,Yarmukhamedov:2011kd}. 
The function $d_{l}(\eta)$ is given by 
\begin{align}
d_{l>0}(\eta)&=\Pi_{j=1}^{l}\left(1+\frac{\eta^2}{j^2}\right)
=\frac{\Gamma(l\!+\!1\!+\!i\eta)\Gamma(l\!+\!1\!-\!i\eta)}
{\Gamma(l\!+\!1)^2\Gamma(1\!+\!i\eta)\Gamma(1\!-\!i\eta)}\nonumber\\
&=\frac{\Gamma(2l+2)^2}{2^{2l}\Gamma(l+1)^2}\,
\frac{C_{l}^2(\eta)}{C_{0}^2(\eta)}\,.
\end{align}

The poles of $S_{l}$ due to physical bound states coincide with the ones of 
${\cal A}_{l}$ as they are related to zeros of the function 
$\displaystyle F_{l}^{-1}=\big[K_{l}-p^{2l+1}d_{l}(\eta_p)2\eta_p H\big]$. 
Close to the bound-state momentum $p\sim i\gamma$, $S_{l}$ behaves 
as~\cite{Mukhamedzhanov:1999zz,Sparenberg:2009rv}
\begin{align}
S_{l}(p\sim i\gamma)&\sim (-1)^{l+1}i e^{i\pi\eta_{\gamma}}
\frac{|C_b|^2}{p-i\gamma}\nonumber\\
&=
(-1)^{l}e^{i\pi\eta_{\gamma}}\frac{\gamma}{\mu}
\frac{|C_b|^2}{(E+B)}\,,
\label{eq:Spole}
\end{align}
with $B=\gamma^2/(2\mu)$, $\eta_\gamma=-i k_C/\gamma$ 
(Eq.(13) of Ref.~\cite{Sparenberg:2009rv} has a misprint, see Ref.~\cite{Mukhamedzhanov:1999zz}), 
and $C_b$ the ANC of the bound-state 
wave function. 

The residues of ${\cal A}_{l}$ and $S_{l}$ are related via 
\begin{align}
{\rm Res}{\cal A}_{l}
=\frac{\pi}{\mu\gamma}{\rm Res}S_{l}
=(-1)^{l}e^{i\pi\eta_\gamma}\frac{\pi}{\mu^2}|C_{b}|^2\,.
\label{eq:anc}
\end{align}
From Eqs.(\ref{eq:Spole}) and (\ref{eq:anc}) one arrives at 
\begin{align}
|C_b|^2={}&-2\mu\gamma^{2l}
\left[\frac{\Gamma(l\!+\!1+\!\eta_\gamma)}{\Gamma(l\!+\!1)}\right]^2\nonumber\\
{}&\times\left\{\frac{d}{dE}\left[F_{l}(p^2)^{-1}\right]_{E=-B}\right\}^{-1}
\,.
\label{eq:anc2}
\end{align}
The relation between the general form of ${\cal A}_{l}$ and the sum of 
on-shell Feynman diagrams derived from EFT is given by 
\begin{align}
{\cal A}_{l}(p^2)={}&\left[p^{l}e^{-\pi\eta_p/2}\Gamma(l\!+\!1+\!i\eta_p)\right]^2
\left[-\frac{2\pi}{\mu}\frac{F_{l}(p^2)}{\Gamma(l\!+\!1)^2}\right]
\nonumber\\
={}&\left[\left(\frac{p}{\mu}\right)^{l}e^{-\pi\eta_p/2}
\Gamma(l\!+\!1+\!i\eta_p)\right]^2
\nonumber\\
{}&\times\left[\frac{2\pi}{\mu}\,
{\cal D}^{(l)}(E,0)\right]
\,,
\label{eq:FDrel}
\end{align}
where ${\mathcal D}^{(l)}(p_0,\bm p)$ is the dressed dimer 
propagator. 
Comparing Eqs.(\ref{eq:FDrel}) and (\ref{eq:anc2}) one gets 
\begin{align}
|C_b|^2={}&\frac{\gamma^{2l}}{\pi\mu^{2l-2}}\left[\Gamma(l\!+\!1+\!\eta_b)\right]^2
\nonumber\\
{}&\times\frac{2\pi}{\mu}
\left\{\frac{d}{dE}\left[{\cal D}^{(l)}(E,0)\right]^{-1}\Big|_{E=-B}\right\}^{-1}
\nonumber\\
={}&\frac{\gamma^{2l}}{\pi\mu^{2l-2}}\left[\Gamma(l\!+\!1+\!\eta_b)\right]^2
\frac{2\pi}{\mu}{\cal Z}\,.\label{eq:ANCs}
\end{align}
This relation between the ANC $C_b$ and the wave function renormalization constant ${\cal Z}$ is valid for any partial wave, up to a convention-dependent multiplicative factor. 
In EFT, different ways of defining the interaction couplings lead to different multiplicative factors for the dressed dimer propagator and, consequently, ${\cal Z}$. This affects the relation between the latter and the scattering amplitude without any physical implication~\cite{Griesshammer:2004pe}. Here we follow the same convention used in Refs.~\cite{Higa:2020kfs,Griesshammer:2004pe}.

\subsection{Bayesian Analysis}
\label{sec:Bayesian}

In this subsection we list the relations essential for drawing Bayesian inferences. If we represent the parameters of the theory by the vector 
$\bm \theta$ and the data set by $D$, then the probability for the parameters to be true given the data and some proposition $H$ is 
the posterior probability distribution 
\begin{align}\label{eq:Posterior}
    P(\bm\theta|D,H)=\frac{ P(D|\bm\theta,H)P(\bm\theta|H)}{P(D|H)}\, ,
\end{align}
where $P(\bm\theta|H)$ is the prior distribution of the parameters based on assumptions made in the proposition $H$. EFT power counting estimates about the  sizes of couplings and parameters enter in the construction of  $P(\bm\theta|H)$. 
The likelihood function is defined as
\begin{align}
P(D|\bm{\theta},H)={}&\prod_{i=1}^N \frac{1}{\sigma_i\sqrt{2\pi}}
\exp\left\{-\frac{[y_i-\mu_i(\bm\theta)]^2}{2\sigma_i^2}\right\}\nonumber\\
={}&\exp\left\{-\sum_{i=1}^N\frac{[y_i-\mu_i(\bm\theta)]^2}{2\sigma_i^2}\right\}\prod_{i=1}^N \frac{1}{\sigma_i\sqrt{2\pi}}\nonumber\\
\equiv {}& e^{-\chi^2/2} \prod_{i=1}^N \frac{1}{\sigma_i\sqrt{2\pi}}\, ,
\end{align}  
if the data set $D$ has $N$ results $y_i$ with measurement errors $\sigma_i$. $\mu_i(\bm\theta)$ are the theory predictions.  The likelihood is maximized by minimizing the $\chi^2$. The \emph{evidence} $P(D|H)$ is obtained from the marginalization of the likelihood over the parameters as
\begin{align}\label{eq:evidence}
P(D|H)= \int \prod_{i}d\theta_i  P(D|\bm\theta,H)P(\bm\theta|H)\, ,
\end{align}
which ensures that we have a normalized posterior distribution: 
$\int \prod_i d\theta_i P(\bm\theta|D,H) =1$.

Parameter estimation does not require calculation of the evidence, an overall normalization to the posterior,  in Eq.~(\ref{eq:Posterior}). Markov Chain Monte Carlo (MCMC) algorithms such as Metropolis-Hastings can be used to generate $P(D|\bm\theta,H)P(\bm\theta|H)$. Given the posterior we can estimate functions of the parameters $f(\bm p; \bm\theta)$ as
\begin{align}
    \langle f(\bm{p})\rangle&= \int \prod_i d\theta_i f(\bm{p};\bm\theta) 
    P(\bm\theta|D,H)\, .
\end{align}
Similarly the posterior of a single parameter $\theta_i$ is  
\begin{align}
    P(\theta_i|D,H)=\int \prod_{j\neq i}d\theta_j P(\bm\theta|D,H)\, .
\end{align}

The evidence is needed when we want to compare theories. Suppose we have two theories $M_A$ and $M_B$, then we can quantify the relative support for the theories by the data as
\begin{align}
    \frac{P(M_A|D,H)}{P(M_B|D,H)}=\frac{P(D|M_A,H)}{P(D|M_B,H)}\times \frac{P(M_A|H)}{P(M_B|H)}\, ,
\end{align}
where the first factor on the right hand side of the equation is the ratio of evidences of the corresponding theories and the second factor is the prior odds.  We set the prior odds to one if the theories we compare are expected to reproduce the data with similar accuracy with the given prior knowledge. In our analysis the two EFTs we compare have the exact same momentum dependence up to NLO, and at NNLO they receive similar sized corrections so we expect the prior odds to be 1. This is in contrast to the comparison done in Ref.~\cite{Premarathna:2019tup} where the two theories had different accuracy at LO and the higher order corrections moved the EFTs closer to data at different rates. The current theory comparison is simpler. 

We calculate the evidences using Nested Sampling (NS)~\cite{Skilling:2006}. There are various implementations of NS, see review article by Brewer~\cite{Brewer:2018}. We use MultiNest~\cite{Feroz:2009} implemented in Python~\cite{Nestle} and Diffusive Nested Sampling~\cite{Brewer:2011} to compute the evidences which also generates the posterior distribution for the parameters.

The \pBe ~calculation presented here is very similar to the \nLi ~one in Ref.~\cite{Higa:2020kfs} once the Coulomb interactions are added using the formalism developed for ~\HeAlpha~\cite{Higa:2016igc}. There is one crucial difference between the \nLi ~and \pBe ~calculations. In the former, the energy of the excited $^7\mathrm{Li}^\star$ core is smaller than the binding energy of the $^8$Li ground state. This in principle requires that the excited core be included explicitly in the \nLi ~calculation, though in practice the momentum dependence in \EFTgsHead{} and \EFTstarHead{} are the same up to NNLO corrections. In the latter \pBe ~calculation,  the energy of the excited $^7\mathrm{Be}^\star$ core $E_\star\sim \SI{429}{\kilo\eV}$ is larger than the binding energy of the $^8$B ground state $B\sim\SI{136}{\kilo\eV}$. Thus at energies $E\sim B\ll E_\star$ below the excitation of the $^7\mathrm{Be}^\star$ core, 
one expects that \EFTgsHead{}  that doesn't include the $^7\mathrm{Be}^\star$ core explicitly might be sufficient. 
At higher energies $E\gtrsim E_\star$ such as to describe the resonance M1 capture, one should include the $^7\mathrm{Be}^\star$ explicitly as the resonance energy is above the excitation energy of $^7\mathrm{Be}^\star$.
We restate that the expressions for \pBe{} E1 transition in \EFTgsHead{} and \EFTstarHead{} have the same momentum dependence up to NLO, though, the latter contains explicit $^7\mathrm{Be}^\star$ contributions in the ANCs as shown later in Eq.~\ref{eq:corsssectionStar}. We use Bayesian evidences to compare these two EFTs over a range of capture energies to quantify the contribution of the excited $^7\text{Be}^\star$ core.

Having described the general framework for the Coulomb and strong interaction, it is appropriate to discuss how our results differ from previous EFT calculations~\cite{Zhang:2013kja,Zhang:2014zsa,Ryberg:2014exa,Zhang:2015ajn,Zhang:2017yqc} in the treatment of the short distance physics. The differences in the EFT power countings is discussed later in Sections~\ref{sec:Analysis} and \ref{sec:Conclusions}. Here we briefly explain the differences in the bound and scattering state construction, independent of the power counting.
A more detailed comparison is given in Sect.~3.1 of Ref.~\cite{Higa:2020kfs}.

Previous EFT calculations consider mixing in the $^5P_2$, $^3P_2$ and $^3P_2^\star$ bound state $p$-wave channels. However, this mixing is implemented using a single auxiliary dimer field unlike the coupled channels construction presented here. Once this single auxiliary dimer field is integrated out of the theory,  it generates proton-core interactions that are not the most general set allowed by low energy symmetries. For example, the mixing in $^3P_2$-$^3P_2^\star$ can be expressed in terms of the couplings $h_{(^3P_2)}$ and $h_{(^3P_2^\star)}$ (using notation from Ref.~\cite{Zhang:2017yqc})   of the elastic $^3P_2$ and $^3P_2^\star$ channels, respectively.  Such modeling of short-distance physics without any underlying symmetry  
is precisely what one would like to avoid in an EFT. Previous EFT calculations describe the wave function renormalization constant involving mixing between three separate channels in terms of a single effective momenta $r_1$. However, the amplitude mixing three channels would in general contain six independent matrix elements (due to Hermiticity) that would correspond to six scattering volumes and six effective momenta at low energy. This would require 6 effective momenta to describe the wave function renormalization constants. Further, the renormalization condition in earlier calculations, for instance Eq.~(46) of Ref.~\cite{Zhang:2017yqc}, ignores mixing in the amplitudes allowed by the interactions, such as factors of $h_{(^5P_2)}h_{(^3P_2)}$ in the external states from Table II of Ref.~\cite{Zhang:2017yqc}. These mixing terms would leave the wave function renormalization constants RG scale dependent. 
In contrast, in the coupled-channels calculation we present, we mix only the  $^3P_2$ and $^3P_2^\star$ bound state channels, and write the RG scale invariant wave function renormalization constant using 3 effective momenta. The formalism presented here allows for a $^5P_2$-$^3P_2$-$^3P_2^\star$ coupled-channels calculation. However, as discussed earlier, the rational for treating $^5P_2$ as a single channel at LO is physically motivated by the large branching ratio of the $S=2$ spin channel.

In the $s$-wave scattering state calculation, 
Zhang \emph{et al.} in their latest work ~\cite{Zhang:2017yqc} model the proton-core short distance interaction using a single dimer field. This makes the $^3S_1$-$^3S_1^\star$ coupling to be the geometric mean of the couplings in the $^3S_1$ and $^3S_1^\star$ channels i.e. scattering length $a_{11}$ becomes a function of $a_{12}$ and $a_{22}$ though there is no underlying symmetry to expect such a result.

 \section{Capture calculation}
 \label{sec:Capture}

\begin{figure}[tbh]
\begin{center}
\includegraphics[width=0.35\textwidth,clip=true]{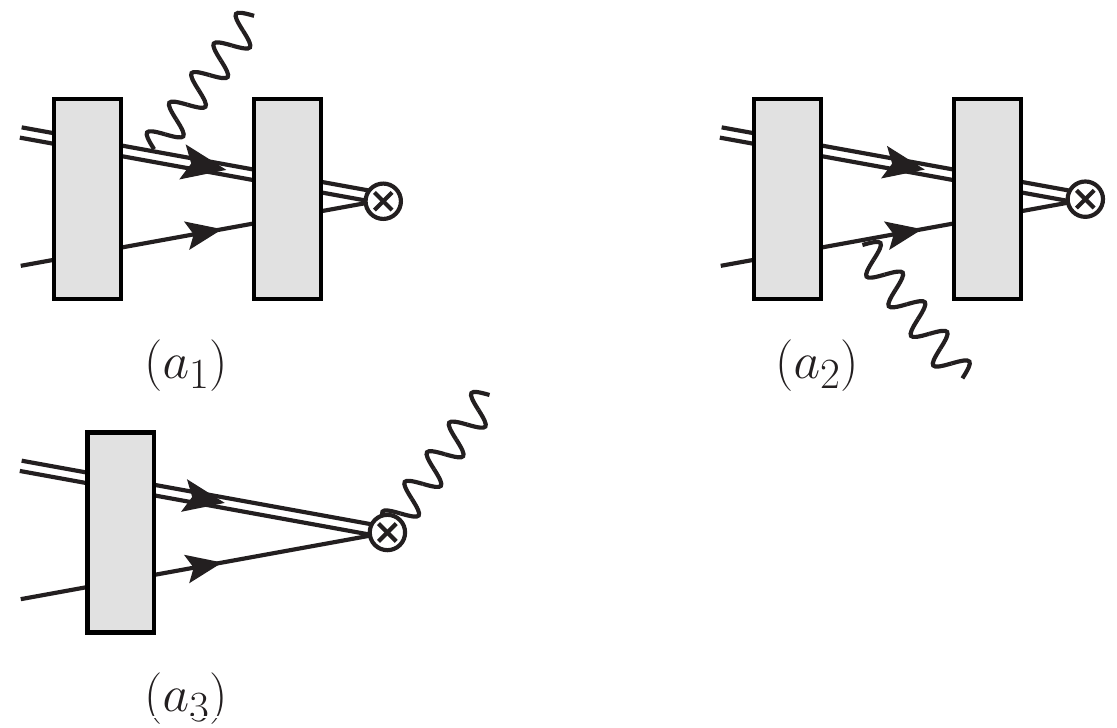}
\end{center}
\caption{\protect E1 capture without initial state strong interactions. Double solid line represents a $^7$Be or $^7\mathrm{Be}^\star$ 
core as appropriate, single solid line a proton, wavy line a photon, and $\otimes$ represents the final $^8$B bound state. The gray rectangles represent Coulomb interactions between the charged particles. The photons are minimally coupled to the charged particles.}
\label{fig:DiagramA}
\end{figure}

\begin{figure}[tbh]
\begin{center}
\includegraphics[width=0.47\textwidth,clip=true]{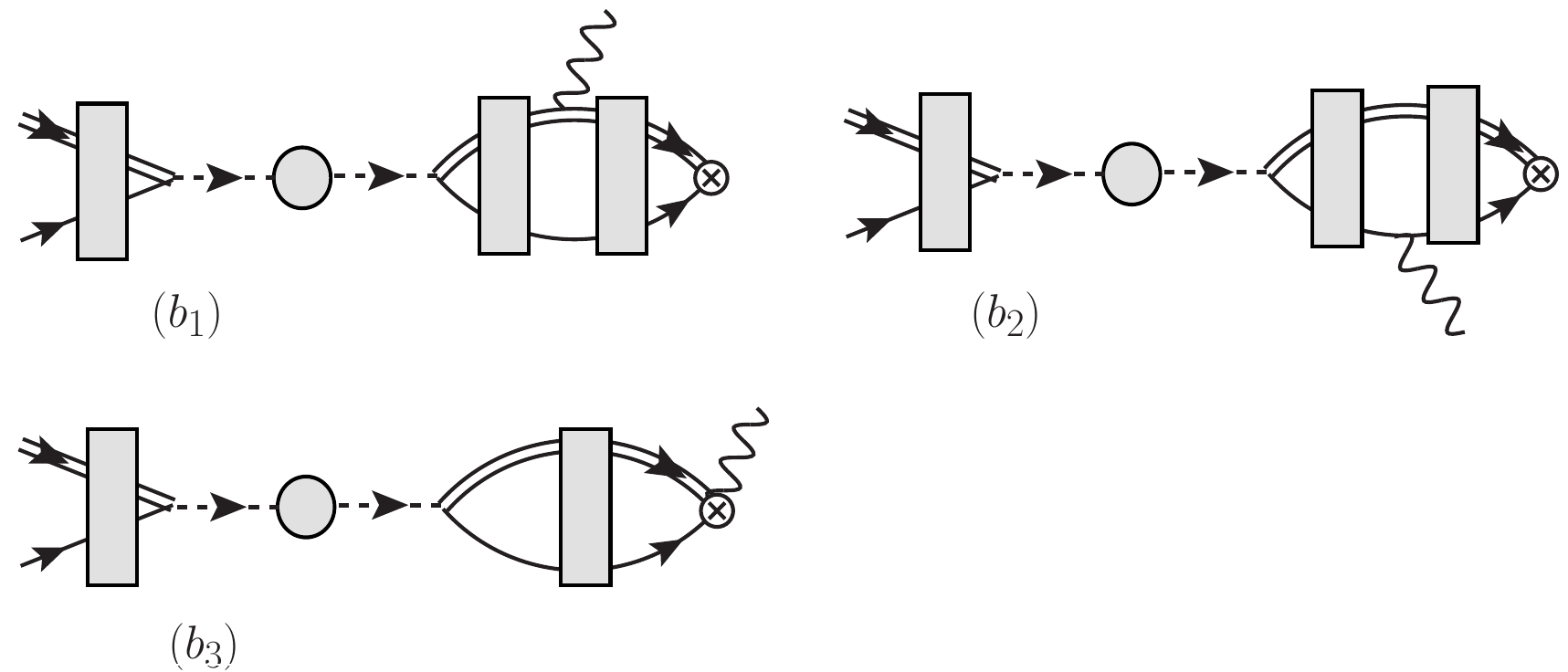}
\end{center}
\caption{\protect E1 capture with initial state strong interactions.
The dashed line with the gray blog represents initial state $s$-wave strong and Coulomb interactions. The rest of the notation is the same as in Fig.~\ref{fig:DiagramA}.}
\label{fig:DiagramB}
\end{figure}

\begin{figure}[tbh]
\begin{center}
\includegraphics[width=0.47\textwidth,clip=true]{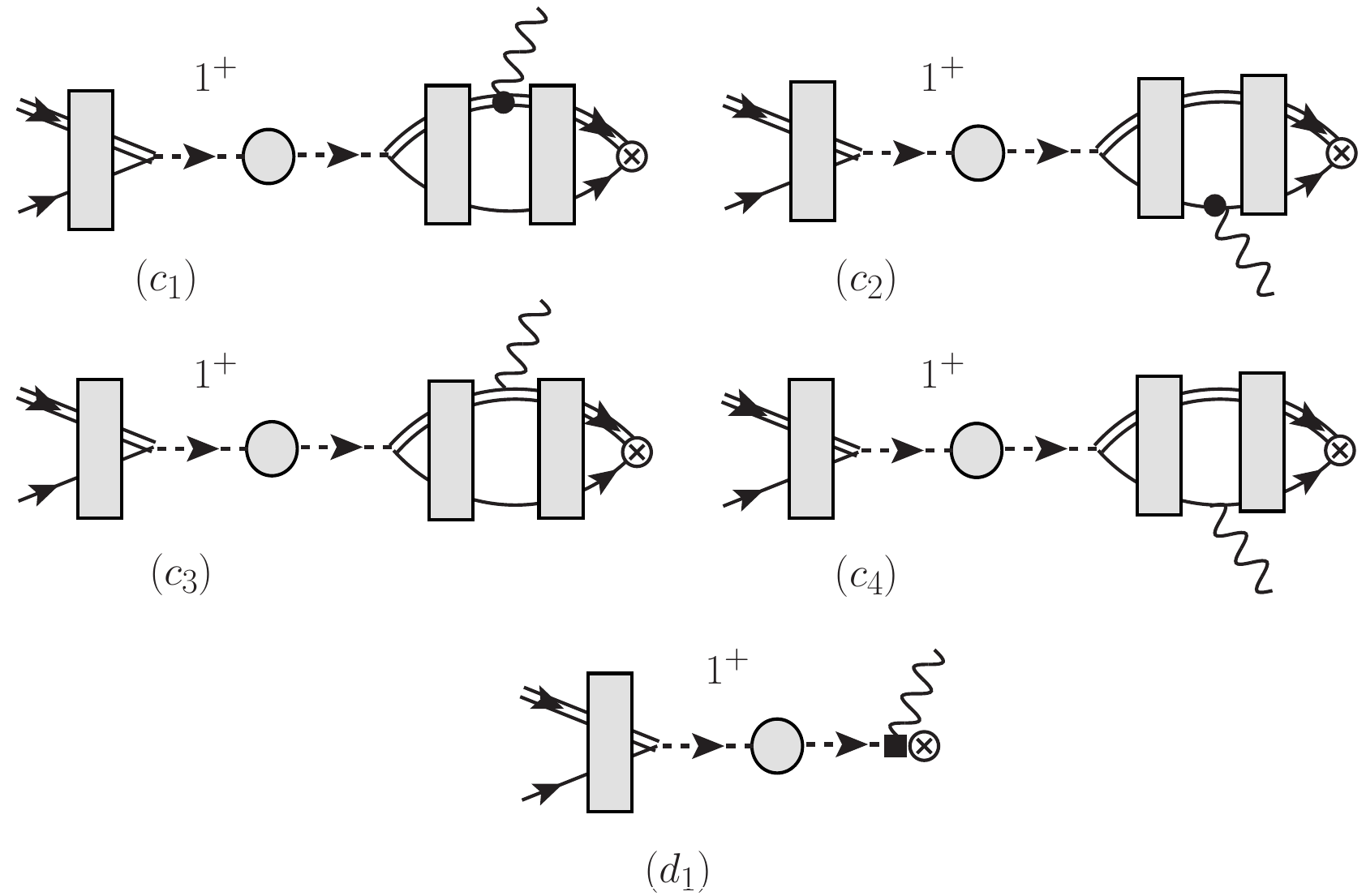}
\end{center}
\caption{\protect M1 capture through the resonant $1^+$ state. The dashed line with the gray blog represents $p$-wave interaction in the $1^+$ resonance channel, {\Large\textbullet} represents magnetic photon coupling and  $\blacksquare$ the two-body current. The rest of the notation is the same as in Fig.~\ref{fig:DiagramA}.}
\label{fig:M1capture}
\end{figure}

The E1 capture to the $^8$B ground state is given by the diagrams in Figs.~\ref{fig:DiagramA}, ~\ref{fig:DiagramB}. The excited $^7\mathrm{Be}^\star$ core contributes only in the spin $S=1$ channel, and only  in Fig.~\ref{fig:DiagramB}. The $S=2$ channel contribution in \EFTgsHead{} and \EFTstarHead{} has the same expression. 

The E1 squared amplitude for capture in the $S=2$ channel can be written as
\begin{align}\label{eq:E1amplitude}
    |\mathcal M_\mathrm{E1}^{(^5P_2)}|^2={}&(2j+1)\left(\frac{Z_c m_p}{M}-\frac{Z_p m_c}{M}\right)^2\frac{64\pi\alpha M^2}{\mu}\nonumber\\
    &
    \Bigg[ 
    \left|X(p)- \frac{\mathcal A_0(a_0^{(2)},p)}{C_0(\eta_p)}[B(p,\gamma)+J_0(-ip)]\right|^2
   \nonumber\\
   &+2|Y(p)|^2\Bigg] \frac{2\pi}{\mu}\mathcal Z^{(^5P_2)}\, ,
\end{align}
where $j=2$ and the $s$-wave contribution  from  Fig.~\ref{fig:DiagramA} is given by
\begin{multline}
    X(p)=C_0(\eta_p)+\frac{2\gamma}{3}\Gamma(2+k_C/\gamma) 
 \int_0^\infty d\,r\, r W_{-\frac{k_C}{\gamma},\frac{3}{2}}(2\gamma r)\\
 \times\frac{\partial}{\partial r}\frac{F_0(\eta_p,p r)}{p r}\, ,
\end{multline}
whereas the $d$-wave contribution is 
\begin{multline}
    Y(p)=\frac{2\gamma}{3}\Gamma(2+k_C/\gamma)
    \int_0^\infty d\,r\, r W_{-\frac{k_C}{\gamma},\frac{3}{2}}(2\gamma r)\\
 \times\left(\frac{\partial}{\partial r}+\frac{3}{r}\right)\frac{F_2(\eta_p,p r)}{p r}\, .
\end{multline}
The regular Coulomb wave functions are
\begin{align}
    F_l(\eta_p,\rho)= C_l(\eta_p) 2^{-l-1}(-i)^{l+1} M_{i\eta_p, l+1/2}(i2\rho)\, ,
\end{align}
with conventionally defined Whittaker functions $M_{k,\mu}(z)$ and $W_{k,\mu}(z)$.
The initial state strong interactions in Fig.~\ref{fig:DiagramB} are given by the $s$-wave amplitude
\begin{align}\label{eq:swaveA}
    \mathcal A_0(a_0^{(2)},p)=\frac{2\pi}{\mu}\frac{[C_0(\eta_p)]^2}{-\frac{1}{a_0^{(2)}}
    -2 k_CH(\eta_p)}\, ,
\end{align}
with the scattering length $a_0^{(2)}=-3.18^{+0.55}_{-0.50}\, \si{\femto\meter}$~\cite{Paneru:2019jyz}. 
The loop contribution from $(b_1)$, $(b_2)$ in Fig.~\ref{fig:DiagramB} is
$B(p,\gamma)$ 
and the contribution from 
$(b_3)$ is $J_0(-ip)$. The linear combination 
$B(p,\gamma)+J_0(-ip)$ is finite, and it is presented in  Appendix~\ref{sec:Integrals}. 
The wave function renormalization constant in the $^5P_2$ channel is 
 \begin{align}\label{eq:Zphi}
     \frac{2\pi}{\mu}\mathcal Z^{(^5P_2)}
    ={}& -{2\pi}\Bigg[r_1^{(^5P_2)}
     +i\frac{k_C^2(k_C^2-\gamma^2)}{\gamma^3} 
     H'\left(-i \frac{k_C}{\gamma}\right)\nonumber\\
   {}&  -4 k_C H\left(-i\frac{k_C}{\gamma}\right)\Bigg]^{-1}\nonumber\\
     \approx{}&-\frac{2\pi}{r_1^{(^5P_2)}-\SI{1.5538}{\mega\eV}}\, ,
 \end{align}
 where $r_1^{(^5P_2)}$ is the $p$-wave effective momentum in the $^5P_2$ channel. In the above we expanded the ERE around the binding momentum $\gamma$ instead of $p=0$~\cite{Phillips:1999hh,Higa:2016igc}. 
 In \EFTgsHead{}, the capture from initial $^3S_1$ state to the $2^+$ ground state is given by a similar expression as the above Eq.~(\ref{eq:E1amplitude}) with the replacements $a_0^{(2)}\rightarrow a_0^{(1)}$ for the scattering length in the $^3S_1$ channel,  and $r_1^{(^5P_2)}\rightarrow r_1^{(^3P_2)}$ for the effective momentum in the $^3P_2$ channel.  The total E1 cross section is 
 \begin{align}\label{eq:crosssection}
     \sigma_\mathrm{E1}(p)=\frac{1}{16\pi M^2}\frac{k_0}{p}\frac{1}{8}&\left[|a|^2 |\mathcal M^{(^5P_2)}|^2\right.\nonumber\\{}&\left.+(1-|a|^2)|\mathcal M^{(^3P_2)}|^2\right]\, , 
 \end{align}
 where $k_0=(\gamma^2+p^2)/(2\mu)$ is the radiative photon c.m. energy. 
 Setting the Clebsch-Gordan coefficient $a=1/\sqrt{2}$ describes the $2^+$ $^8$B ground state as a $p_{3/2}$ proton state~\cite{Grigorenko:1999mp}.

 The M1 capture cross section from Fig.~\ref{fig:M1capture} is given by
 \begin{align}\label{eq:sigmaM1}
     \sigma_\mathrm{M1}(p)={}&\frac{1}{16\pi M^2}\frac{k_0}{p}\frac{1}{8}
     |b|^2 (2j+1)\frac{k^2\mu^3}{p^2} \frac{8\pi\alpha_e M^2}{m_p^2}\nonumber\\
     {}&\times\frac{|\mathcal A_1(p)|^2}{[C_1(\eta_p)]^2}\frac{2\pi}{\mu}\mathcal Z^{(^5P_2)}
      \frac{\mu^2}{432\pi^4}|\xbar{L}_{22}(p)|^2\, ,\nonumber\\
       \xbar{L}_{22}(p)\equiv{}&\frac{2\pi}{\mu} \left\{\frac{9\pi}{\sqrt{40}} 
       \left[3 g_c + g_p+4\mu m_p\left(\frac{Z_c}{m_c^2}+\frac{Z_p}{m_p^2}\right)\right]\right.\nonumber\\
       {}&\times\left.\frac{\overline{C}(p)}{\mu^2} -\beta_{22}\right\}\, ,
 \end{align}
 where the divergence in the integral $C(p)$ is regulated by the two-body current coupling $L_{22}\sim - 2\pi\beta_{22}/\mu$
 that subtracts the divergent pieces up to an overall renormalization constant $\beta_{22}\sim \Lambda$.  
 $\overline C(p)$ is the regulated $C(p)$ without the divergences, see Appendix~\ref{sec:Integrals}. The Clebsch-Gordan coeeficient $b=\sqrt{5/6}$ corresponds to making the $1^+$ resonance a $p_{1/2}$ proton state~\cite{Grigorenko:1999mp}. The M1 contribution is peaked near the momentum $p\sim p_R$ and we expect the renormalized combination $|\xbar{L}_{22}(p_R)|\sim1$ from the power counting.
 
 The $^5P_1$ scattering amplitude is 
 \begin{align}\label{eq:A1pwave}
     \mathcal A_1(p)={}&
     \frac{2\pi}{\mu}\frac{e^{i2\sigma_1}}{p\cot\delta_1-i p}\nonumber\\ 
     ={}& \frac{2\pi}{\mu}9 C_1^2(\eta_p)e^{i2\sigma_1} p^2
\left[-\frac{1}{a_1^{(^5P_1)}} + \frac{1}{2}r_1^{(^5P_1)} p^2\right.\nonumber\\
{}&\left.-2 k_c(k_c^2+p^2) H\left(\frac{k_c}{p}\right)\right]^{-1}\, ,
 \end{align}
 where the scattering volume $a_1^{(^5P_1)}$ and effective momentum $r_1^{(^5P_1)}$ are tuned to reproduce the $1^+$ resonance energy 
 $E_R=p_R^2/(2\mu)=\SI{0.630+-0.003}{\mega\eV}$ and width 
 $\Gamma_R=\SI{0.0357+-0.0006}{\mega\eV}$~\cite{Junghans:2003bd} similar to the M1 contribution in \nLi~\cite{Fernando:2011ts}. Near the resonance momentum $p_R$, the $p$-wave phase shift $\delta_1(p)$ in the incoming channel increases rapidly through $\pi/2$ from below. Thus we impose the conditions~\cite{Fernando:2011ts} 
 \begin{align}\label{eq:Resonance}
    \cot\delta_1|_{E=E_r}&=0\, ,\nonumber \\
    \frac{d\cot\delta_1}{dE}|_{E=E_r} &\equiv -\frac{2}{\Gamma_r}<0\, ,
    \end{align}
 which produce a Breit-Wigner form near the resonance
    \begin{align}
    \cot\delta_1(E)&\approx\cot\delta_1(E_r) +(E-E_r)\cot'\delta_1(E_r)\nonumber\\
    &=-\frac{2}{\Gamma_r}(E-E_r)\, ,\nonumber\\
  \frac{2\pi}{\mu}\frac{e^{i2\sigma_1}}{p\cot\delta_1-i p} &\approx   \frac{2\pi}{\mu p_r}\frac{e^{i2\sigma_1}(-\Gamma_r/2)}{E-E_r+i\Gamma_r/2}\, . 
\end{align}
Imposing the conditions in Eq.~(\ref{eq:Resonance}) on the ERE in Eq.~(\ref{eq:A1pwave}) gives $a_1^{(^5P_1)}=\SI{-108.13}{\femto\meter^{3}}$, $r_1^{(^5P_1)}=-111.23$ MeV for the central values.

The $S$-factor in \EFTgsHead{} is  then
 \begin{align}\label{eq:SfactorEFT}
      S_{17}(p)&=\frac{p^2}{2\mu}e^{2\pi\eta_p}
      \left[\sigma_\mathrm{E1}(p)+\sigma_\mathrm{M1}(p)\right]\, .
 \end{align}
 
 In \EFTstarHead{}, the E1 capture in the $S=2$ channel is the same as Eq.~(\ref{eq:E1amplitude}). In the $S=1$ channel, the squared amplitude is slightly modified to
\begin{align}\label{eq:E1amplitudeStar}
    |\mathcal M_\mathrm{E1,\star}^{(^3P_2)}|^2=(2j+1)\left(\frac{Z_c m_p}{M}-\frac{Z_p m_c}{M}\right)^2\frac{64\pi\alpha M^2}{\mu}\nonumber\\
    \times\left[|\mathcal A_\star^{(^3P_2)}(p)|^2+2|Y(p)|^2\right]
    \frac{2\pi}{\mu}\mathcal Z^{(^3P_2)}\, ,
\end{align}
where
\begin{align}
    \mathcal A_\star^{(^3P_2)}(p)=X(p)- \frac{\mathcal A^{(11)}(p)}{C_0(\eta_p)}[B(p,\gamma)+J_0(-ip)]\nonumber\\
    - \frac{\mathcal A^{(12)}(p)}{C_0(\eta_p)}[B(p_\star,\gamma_\star)+J_0(-ip_\star)]
    \frac{\sqrt{\mathcal Z^{(^3P_2^\star)}}}{\sqrt{\mathcal Z^{(^3P_2)}}}\, .
\end{align}
The M1 cross section at this order of the calculation is given by Eq.~(\ref{eq:sigmaM1}). 
Thus the total c.m. cross section and $S$-factor in \EFTstarHead{} are:
 \begin{align}\label{eq:corsssectionStar}
     \sigma_\mathrm{E1,\star}(p)={}&\frac{1}{16\pi M^2}\frac{k_0}{p}\frac{1}{8}
     \left[ |a|^2|\mathcal M^{(^5P_2)}_\mathrm{E1}|^2\right.\nonumber\\
     {}&\left.+(1-|a|^2)|\mathcal M^{(^3P_2)}_\mathrm{E1,\star}|^2\right]\, ,\nonumber\\
     S_{17,\star}(p)={}&\frac{p^2}{2\mu}e^{2\pi\eta_p}
     \left[\sigma_\mathrm{E1,\star}(p)+\sigma_\mathrm{M1}(p)\right]\, ,
 \end{align}
with $|a|^2=1/2$.

 \section{EFT power counting}
 \label{sec:PowerCounting}
 
\begin{figure}[tbh]
\begin{center}
\includegraphics[width=0.47\textwidth,clip=true]{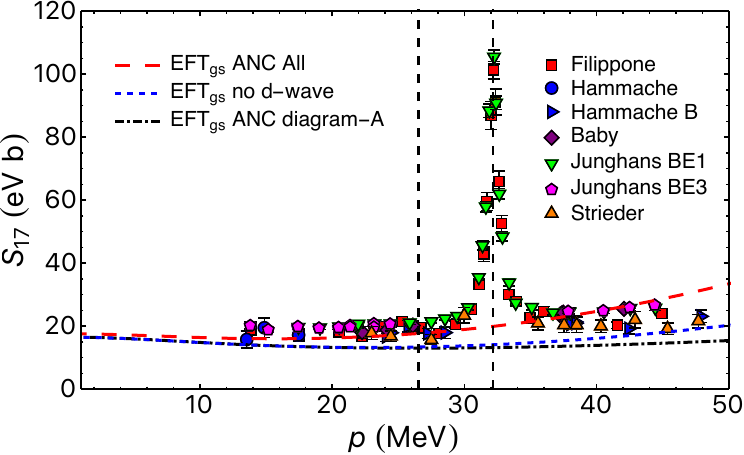}
\end{center}
\caption{\protect $S$-factor for \pBe ~in \EFTgsHead{}. The grid lines are at the $^7\mathrm{Be}^\star$ inelasticity $\gamma_\Delta=26.5$ MeV and the $1^+$ resonance momentum $p_R=32.2$ MeV.  
The dot-dashed (black) curve includes diagrams with only $s$-wave contributions and no initial state strong interactions, 
the short-dashed (blue) curve includes all diagrams except $d$-wave contributions, 
and the long-dashed (red) curve includes full contributions from all diagrams. 
Results were fitted to ANCs as explained in the text.}
\label{fig:Survey}
\end{figure}
 
 At low energy, proton-$^7$Be scattering in the entrance channel in \pBe ~is peripheral due to the Coulomb repulsion. It is expected that the capture to the $^8$B ground state can still proceed without initial-state short-range strong interaction as it is a very loosely bound state with the bound-state wave function extending over a large spatial distance $\sim1/\gamma$.  This is borne out by a direct calculation, in \EFTgsHead{}, from Eq.~(\ref{eq:crosssection}) at  threshold (numerically evaluated at c.m. energy $E_0=50$ eV in this work)
 \begin{multline}\label{eq:SfactorThreshhold}
     S_{17}/C_{1,\zeta}^2\approx 35.6(1
     - a_0\,\SI{0.00266}{\femto\meter^{-1}}\\
    +0.0657+\dots)\, \si{\eV\barn\femto\meter}\, ,
 \end{multline}
in either channel $\zeta={}^5P_2$, $^3P_2$. The first term is the $s$-wave contribution from Fig.~\ref{fig:DiagramA} that contains no initial state strong interaction, the second term is the leading contribution from Fig.~\ref{fig:DiagramB} with $s$-wave scattering length $a_0$ (written in fm units above), and the last term is the $d$-wave contribution from Fig.~\ref{fig:DiagramA}.  These numbers are consistent with the values in Ref.~\cite{Baye:2000} except our strong interaction contribution (second term) is a factor of 2 larger which comes from squaring the amplitude and keeping the linear $a_0$ term. The relative contributions of the three terms above agree with the result from Zhang {\em et al.}~\cite{Zhang:2017yqc} though they disagree with the overall value a little. 
The $d$-wave contribution in Eq.~(\ref{eq:SfactorThreshhold}) is consistent with a NNLO contribution, and increases in size at higher momentum, as we show,  to be counted as NLO. 

In Fig.~\ref{fig:Survey} we use the \EFTgsHead{} expressions for E1 capture from Eq.~(\ref{eq:SfactorEFT}) without assuming any specific power counting. The M1 capture contributes only in a narrow region that will be added later. The E1 cross section depends on $s$-wave scattering lengths $a_0^{(2)}=-3.18^{+0.55}_{-0.50}\, \si{\femto\meter}$, $a_0^{(1)}=17.34^{+1.11}_{-1.33}\, \si{\femto\meter}$~\cite{Paneru:2019jyz}, and $p$-wave effective momenta $r_1^{(^5P_2)}$, $r_1^{(^3P_2)}$ that we constrain from the measured ANCs~\cite{Trache:2003ir,Tabacaru:2006}  using the relation in Eqs.~(\ref{eq:ANCs}), 
(\ref{eq:Zphi}) along with the normalization of the states defined in Eqs.~(\ref{eq:crosssection}), (\ref{eq:SfactorThreshhold}). We use  
$\mathcal Z^{(^5P_2)}=\num{19.6+-0.8}$ [$r_1^{(^5P_2)}=\SI{-40.3+-1.7}{\mega\eV}$], 
$\mathcal Z^{(^3P_2)}=\num{4.6+-0.4}$ [$r_1^{(^3P_2)}=\SI{-177+-16}{\mega\eV}$].
The dot-dashed curve is the $s$-wave contribution from Fig.~\ref{fig:DiagramA}. Adding the initial state interactions, short-dashed curve, from Fig.~\ref{fig:DiagramB} has a small effect. However, the $d$-wave contribution shown by the long-dashed curve is more substantial. A natural-sized $s$-wave effective range $r_0\sim 1/\Lambda$ in either spin channels has a negligible effect.
From Eq.~(\ref{eq:SfactorThreshhold}), we can estimate the branching ratio for capture to the $^5P_2$ channel to be about 80\% that is mostly determined by $\mathcal Z^{(^5P_2)}/\mathcal Z^{(^3P_2)}\sim 4$.
We develop a power counting based on these observations. 

\begin{figure}[tbh]
\begin{center}
\includegraphics[width=0.47\textwidth,clip=true]{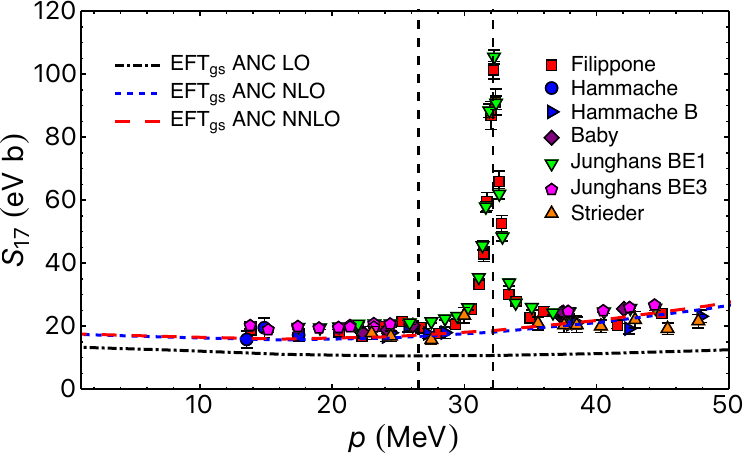}
\end{center}
\caption{\protect $S$-factor for \pBe  ~in \EFTgsHead{}. The grid lines are explained in Fig.~\ref{fig:Survey}. 
The LO, NLO, and NNLO curves are identified in the legends. 
Results were fitted to ANCs in perturbation as explained in the text.}
\label{fig:CaptureEFTNNLO}
\end{figure}

The initial state strong interactions from Fig.~\ref{fig:DiagramB} scale as $2\pi a_0 (B+J_0)/\mu$. For $p\lesssim Q$, the linear combination $2\pi (B+J_0)/\mu$ 
is of order $Q^2/\Lambda$~\cite{Higa:2016igc},
instead of the naive expectation $2\pi (B+J_0)/\mu\sim Q$ that only holds true in the absence of Coulomb interactions~\cite{Higa:2020kfs}. 
In the $S=2$ channel, the $s$-wave scattering length is natural sized $|a_0^{(2)}|\sim \SI{3}{\femto\meter}\sim 1/\Lambda$ making initial state strong interactions $2\pi a_0^{(2)}(B+J_0)/\mu\sim Q^2/\Lambda^2$ a NNLO effect. The larger scattering length in the $S=1$ channel $a_0^{(1)}\sim \SI{17}{\femto\meter}\sim 1/Q$ makes this contribution 
$2\pi a_0^{(2)}(B+J_0)/\mu\sim Q/\Lambda$.  However, the $S=1$ channel branching ratio is around 0.2. 
Thus, all contributions in this channel are one order higher in the perturbation making initial state strong interactions a NNLO contribution in this spin $S=1$ channel as well~\cite{Higa:2016igc}.

The contribution of the $s$-wave effective range $r_0$ in Eq.~(\ref{eq:swaveA}) is also a subleading effect. The effective range contribution $r_0 p^2/2\sim Q^2/\Lambda$ is smaller than the scattering length $1/a_0$ in either spin channels. Moreover we find, similar to $\alpha$-$\alpha$, \HeAlpha, ~\TrAlpha~\cite{Higa:2008dn,Higa:2016igc,Premarathna:2019tup}, cancellations in the linear combination $r_0 p^2/2-2 k_C H(\eta_p)$ suppressing effective range contributions further~\cite{Higa:2008dn}. A general analysis of effective range contributions in shallow systems with Coulomb interactions was done in Refs.~\cite{Schmickler:2019ewl,Luna:2019ufu}.
In our fits we drop the $2 k_C H(\eta_p)$ term in $\mathcal A_0$ of Eq.~(\ref{eq:swaveA}), and similar contributions to $\mathcal A_{11}$, $\mathcal A_{12}$ in Eqs.~(\ref{eq:swaveA11}), (\ref{eq:swaveA12}) that would contribute beyond NNLO.

\begin{figure}[tbh]
\begin{center}
\includegraphics[width=0.47\textwidth,clip=true]{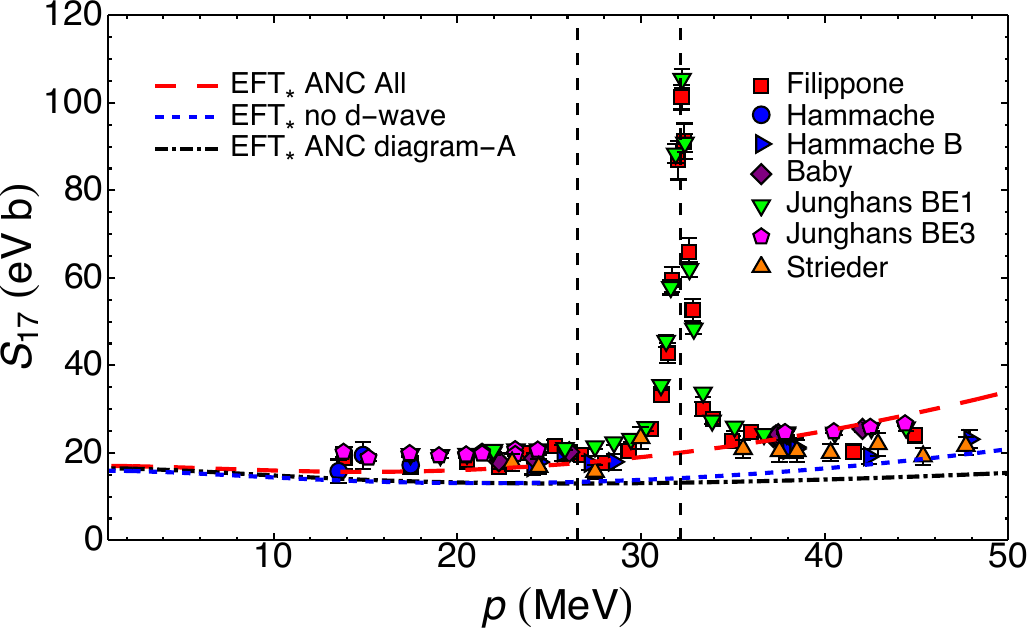}
\end{center}
\caption{\protect $S$-factor for \pBe ~in \EFTstarHead{}. The plots used $r_0^{(1)}=\SI{-30}{\femto\meter}$ and $a_{22}=\SI{-30}{\femto\meter}$ that were varied by 95\% and had negligible effect on the final result. A positive $a_{22}=\SI{30}{\femto\meter}$ also had a negligible effect.
The curves follow the same notation as in Fig.~\ref{fig:Survey}.}
\label{fig:SurveyStar}
\end{figure}

We propose the following power counting in \EFTgsHead{}: The $s$-wave contribution in the $S=2$ channel without initial state strong interaction from Fig.~\ref{fig:DiagramA} constitutes the LO contribution. At NLO, we include the $d$-wave contribution in $S=2$ and the $s$-wave capture in $S=1$ without initial state strong interaction from Fig.~\ref{fig:DiagramA}. Contributions from initial state strong interactions in both the spin 
$S=1$, $2$ channels from Fig.~\ref{fig:DiagramB} enter at NNLO. The $d$-wave contribution in $S=1$ channel contributes at NNLO as well.

The perturbative calculation is shown in Fig.~\ref{fig:CaptureEFTNNLO}. In this calculation we use the zed-parametrization~\cite{Phillips:1999hh} where the exact wave function renormalization is produced at NLO. 
This is achieved by defining the corresponding $p$-wave ERE around the bound-state pole instead of at momentum $p=0$~\cite{Phillips:1999hh,Fernando:2011ts}.
The 20\% jump from LO to NLO even at threshold is due to the inclusion of $S=1$ channel at NLO which is absent entirely at LO in the power counting. We get for the $S$-factor at threshold 13.40(1)(402) \si{\eV\barn}, 17.4(6)(17) \si{\eV\barn}, and 17.6(7)(5) \si{\eV\barn} at LO, NLO, and NNLO, respectively, where the first set of errors are from the input and the second set are the theory errors (with $Q/\Lambda\sim0.3$). The NNLO result compares well with  \SI{18.0+-1.9}{\eV\barn} evaluated in Ref.~\cite{Tabacaru:2006} using the same ANCs.

The nearly identical NLO and NNLO EFT results above can be understood from Eq.~(\ref{eq:SfactorThreshhold}). In the $S=2$ channel, the $s$-wave strong interaction contribution for a natural sized $a_0^{(2)}$ is numerically smaller than the assumed NNLO estimate. The $S=1$ channel initial state $s$-wave strong interaction and the $d$-wave contribution enters at NNLO. However, the large scattering length in this channel $a_0^{(1)}\sim 17$ fm competes and nearly cancels the $d$-wave contribution 
 in Eq.~(\ref{eq:SfactorThreshhold}) with $-a_0^{(1)}\,\SI{0.00266}{\femto\meter^{-1}}+0.0657\approx 0.02$ relative to the leading contribution in this channel which is already a NLO term.  Thus these contributions are  about 
$0.7$\% 
of the LO piece coming from the $S=2$ channel. This behavior persists at higher momenta as well. 

In Fig.~\ref{fig:SurveyStar} we plot the result for \pBe ~in \EFTstarHead{} from Eq.~(\ref{eq:corsssectionStar}) without the resonant M1 contribution. The calculated ANC value $C_{1,^3P_2^\star}^2=\SI{0.1215+-0.0036}{\femto\meter^{-1}}$~\cite{Zhang:2017yqc} was used to determine the wave function normalization 
$\mathcal Z^{(^3P_2^\star)}=\num{7.72+-1.1}$. We use Eq.~(\ref{eq:EFTstarConstrains})  to substitute a $a_{12}>0$ by $r_0^{(1)}$ for convenience here, though a $a_{12}<0$ would work as well. We estimate  $r_0^{(1)}\sim 1/Q\sim a_{22}$ over a large range to generate the curves without assuming any specific power counting. We get results similar to those in Fig.~\ref{fig:Survey}, and we interpret them in a similar manner.

The proposed power counting in \EFTstarHead{} is: The $s$-wave contribution in $S=2$ channel without initial state strong interaction from Fig.~\ref{fig:DiagramA} constitutes the LO contribution. At NLO, we include the $d$-wave contribution in $S=2$ and the $s$-wave capture in $S=1$ without initial state strong interaction from Fig.~\ref{fig:DiagramA}. Contributions from initial 
$s$-wave
strong interactions in both the spin $S=1$, $2$ channels from Fig.~\ref{fig:DiagramB} 
and pure Coulomb initial $d$-wave in $S=1$
 enter at NNLO. At this order, \EFTstarHead{} has three  additional parameters: the wave function renormalization constant $\mathcal Z^{(^3P_2^\star)}$ and two scattering lengths $|a_{22}|\sim |a_{12}|\sim 1/Q$. In addition, when  we apply the \EFTstarHead{} at momenta above the $1^+$ resonance $^8$B state, we include a two-body current 
coupling for the M1 transition.

 \section{Results and Analysis}
 \label{sec:Analysis}

 In this section, we estimate the $S$-factor at threshold $S_{17}(0)$ by constraining the EFT parameters from capture data. We perform the analysis using data at energy $E\leq 500$ keV (momentum  $p\leq28.6$ MeV) that is below the resonance energy $E_R=\SI{630}{\kilo\eV}$ (momentum $p_R=\SI{32.2}{\mega\eV}$). In this low-energy region, named region I, both \EFTgsHead{} and \EFTstarHead{} are applicable and the corresponding fits will be refered to as \EFTgsHead{} I and \EFTstarHead{} I, respectively. We also perform the analysis using capture data up to energy $E\leq 1$ MeV (momentum $p\leq 40.5$ MeV) where we include the M1 contribution to describe the resonance contribution. We call this larger  energy (momentum) region as region II. In region II, only \EFTstarHead{} is applicable that we refer to as \EFTstarHead{} II fits. However, we find the differences between \EFTgsHead{} and \EFTstarHead{} threshold $S$-factors to be very small as the differences are a NNLO effect. 
 This agrees with previous work~\cite{Nunes:1997,Nunes:1999b} that also found the excited core contributions to be small.
 The data, especially at higher momentum,  though, favors \EFTstarHead{} in the evidence calculation.  
 
  In \EFTgsHead{}, the fit parameters are $r_1^{(^5P_2)}$, $r_1^{(^3P_2)}$ in region I. In \EFTstarHead{}, the fit parameters are $r_1^{(^5P_2)}$, $\mathcal Z^{(^3P_2)}$, $\mathcal Z^{(^3P_2^\star)}$, $a_{22}$, $a_{12}$
in region I, and also $\beta_{22}$ in region II for the M1 resonance capture. 
 In the analysis we use data from Filippone {\em et al.}~\cite{Filippone:1984us}, Hammache {\em et al.}~\cite{Hammache:2001,Hammache:1998}, Baby {\em et al.}~\cite{Baby:2003}, Junghans {\em et al.}~\cite{Junghans:2010}, and Strieder {\em et al.}~\cite{Strieder:2001}. We draw Bayesian inferences for the $S$-factor using the capture data. 
 
 An advantage of the Bayesian method over $\chi^2$ fits  is that it allows a natural framework to impose EFT power counting  estimates of the parameters in the data fitting procedure. Details of the method we use are in Ref.~\cite{Premarathna:2019tup}. In the review by Adelberger {\em et al.}~\cite{Adelberger:2010qa}, the $\chi^2$ fit of the theory curve by Descouvemont~\cite{Descouvemont:2004} to data necessitated  inflating the measurement errors to get a smaller reduced $\chi^2$ value with a significant p-value because the global data sets from different experiments were not compatible.  
 In the Bayesian framework, the overall normalization of data can be estimated using the published common-mode-error in a straightforward manner~\cite{Zhang:2014zsa,Zhang:2015ajn,Zhang:2017yqc,Premarathna:2019tup}.

 In the Bayesian fits, we multiply the data from the same measurements by a scale factor $s$ that is drawn from a prior distributed normally with a mean $\langle s\rangle=1$ and standard deviation s.d., following the treatment of systematic errors in Chapter 6 of Ref.~\cite{AGostini:2003}. The s.d.s are set to the published common-mode-errors, see also  Refs.~\cite{Cyburt:2004,Cyburt:2005, Zhang:2015ajn,Zhang:2017yqc}.
 We use for data from Filippone {\em et al.}~\cite{Filippone:1984us} s.d. = 11.9/100, Hammache {\em et al.}~\cite{Hammache:2001} s.d. = 9.0/100, Hammache {\em et al.}~\cite{Hammache:1998} s.d. = 5.0/100, Baby {\em et al.}~\cite{Baby:2003} s.d. = 2.2/100, Junghans {\em et al.}~\cite{Junghans:2010}  s.d. = 2.7/100 for set BE1 and  s.d. = 2.3/100 for set BE3,  Strieder {\em et al.}~\cite{Strieder:2001}  s.d. = 8.3/100, and we label the scaling factors $s_i$ for $i=1,2,\dots,7$ in that order, respectively. These values agree  with Cyburt {\em et al.} in Ref.~\cite{Cyburt:2004, Cyburt:2005} except for the Hammache {\em et al.}~\cite{Hammache:2001} data set where they use a 12.2\% error in the analysis. Zhang {\em et al.}~\cite{Zhang:2015ajn} use similar errors in their Bayesian analysis except they do not assign common-mode-error to the Hammache {\em et al.}~\cite{Hammache:2001} data set, and do not use the Strieder {\em et al.}~\cite{Strieder:2001} data.  The particular value of s.d. that we use in the prior distribution does not affect the quality of the fit as long as it is comparable with the estimates from the experiments. We get the same threshold $S$-factor within the errors from the fits.

 We draw the effective momentum $r_1^{(^5P_2)}$ from an uniform prior $U(\SI{-100}{\mega\eV},\SI{1.5}{\mega\eV})$ consistent with the power counting estimate and the requirement $\mathcal Z^{(^5P_2)}\geq 0$. We know from the \nLi ~calculation that there is a strong correlation between the wave function renormalization constants in the spin $S=2$ and $S=1$ channels~\cite{Higa:2020kfs}. For example, in the EFT without explicit $^7\mathrm{Li}^\star$ core, a single parameter family of effective momenta 
 $r_1^{(^5P_2)}$, $r_1^{(^3P_2)}$ describes the capture cross section. We find a similar behavior in \pBe ~where the capture data is not sensitive to the individual $p$-wave effective momenta and the corresponding wave function renormalization constants. Therefore we fix the wave function renormalization constants in the spin $S=1$ channel by drawing them from a normal distribution as determined 
 by the mean experimental~\cite{Trache:2003ir} and calculated 
 ANCs~\cite{Zhang:2017yqc} (and their errors). Alternatively, constraining the wave function renormalization constants from the known ANC ratios, say taking $\mathcal Z^{(^3P_2)}=\mathcal Z^{(^5P_2)} C_{1,^3P_2}^2/C_{1,^5P_2}^2$, gives similar fits. Note that we impose these constraints so that the $S=1, 2$ spin channels have the hierarchy assumed in the construction of the EFT power counting. The theory expressions themselves can fit the capture data just as well without these EFT assumptions but then one cannot reliably apply the power counting estimates.

\begin{figure}[tbh]
\begin{center}
\includegraphics[width=0.47\textwidth,clip=true]{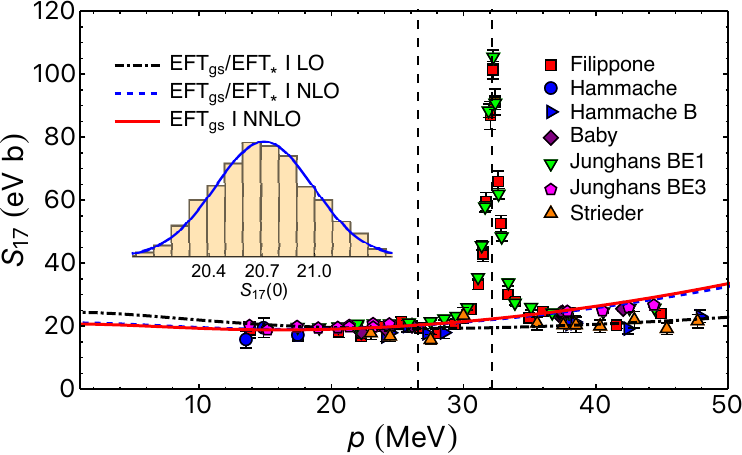}
\end{center}
\caption{\protect $S$-factor for \pBe ~in \EFTgsHead{} fitted  in the momentum region I, $p\leq28.6$ MeV ($E\leq \SI{500}{\kilo\eV}$).
The LO, NLO, and NNLO curves are identified in the  legends.
At LO and NLO, EFT I and \EFTstarHead{} I have the same momentum dependence. The inset shows the posterior distribution for $S_{17}(0)$ in \si{\eV\barn} units at NNLO.}
\label{fig:E1CaptureEFTNNLO}
\end{figure}

  In \EFTstarHead{} we also fit $a_{22}\sim U(\SI{-50}{\femto\meter},\SI{50}{\femto\meter})$ and $a_{12}\sim U(\SI{-50}{\femto\meter},\SI{50}{\femto\meter})$ at NNLO.  For the M1 capture we choose the prior $\beta_{22}\sim U(\SI{-500}{\mega\eV},0)$.
  Near the narrow resonance momentum $p\sim p_R$, $\xbar{L}_{22}(p)$ 
  in Eq.~(\ref{eq:sigmaM1}) is nearly constant, and both positive and negative values for $\beta_{22}$ are compatible.
  We find some evidence from the capture data for a negative $\beta_{22}$ that goes into our prior selection but $\beta_{22}<0$ is not crucial for the analysis. The range of the priors for the EFT couplings were chosen wide enough to accommodate the power counting estimates $r_1^{(^5P_2)}\sim \Lambda$, 
$a_{22}\sim 1/Q$, $a_{11}\sim 1/Q$
and $\beta_{22}\sim \Lambda$.  
We notice that if one restricts $|r_0^{(1)}|\sim 1/Q$, then  $a_{22}$, $a_{12}$ values are further constrained. In our fits, we let the capture data determine $a_{22}$, $a_{12}$ without this constraint to explore the entire parameter space (given by the uniform priors). This doesn't violate the power counting estimate. However, in future one could impose the additional EFT assumption $|r_0^{(1)}|\sim 1/Q$ to derive more  restrictive priors for  $a_{22}$, $a_{12}$.
 
 We note that the normally distributed priors for the scale factors $s_i$s,  chosen for convenience,  are technically incorrect in that they allow for negative values for positive parameters. However, the cumulative prior probability for negative values is much smaller than a percent. For example, a 30\% s.d. translates to a prior cumulative probability of about 0.04\% for a negative value of a positive parameter. {\em A posteriori} the fits give expected values for the parameters and never pick the unphysical negative values in Table~\ref{table:B8norms}. We also verified the robustness of our fits by using a prior for $r_1^{(^5P_2)}$ normally distributed around the prediction from ANC~\cite{Trache:2003ir,Tabacaru:2006} with a 30\% error. The sensitivity of the fits on the priors is found to be insignificant compared to the measurement and theory errors.

\begin{table*}[tbh]
\centering
\caption{The median and the interval containing 68\% of the posterior of the EFT parameters. For comparison, the expected sizes of parameters from the ANCs are:
$\mathcal Z^{(^5P_2)}=\num{19.6+-0.8}$,
$\mathcal Z^{(^3P_2)}=\num{4.6+-0.4}$,
$\mathcal Z^{(^3P_2^\star)}=\num{7.7+-1.1}$.
$\xbar{L}_{22}$ is evaluated at $p=p_R$.}
\begin{ruledtabular}
\begin{tabular}{lrrrrrr}
Theory & $\mathcal Z^{(^5P_2)}$  & $\mathcal Z^{(^3P_2)}$ & $\mathcal Z^{(^3P_2^\star)}$ 
& $a_{22}(\si{\femto\meter})$ & $a_{12}(\si{\femto\meter})$
& $|\xbar{L}_{22}|$\\ \hline
\begin{filecontents*}{paramsTable.csv}
Theory,ancA,dancAup,dancAdown,ancB,dancBup,dancBdown,ancC,dancCup,dancCdown,abb,dabbup,dabbdown,rab,drabup,drabdown,betabb,dbetaup,dbetadown
EFT$_\text{gs}$/EFT$_\star$ I LO,35.8,0.5,0.5,9999,5555,5555,9999,5555,5555,9999,5555,5555,9999,5555,5555,9999,5555,5555
EFT$_\text{gs}$/EFT$_\star$ I NLO,24.6,0.5,0.5,4.6,0.4,0.4,9999,5555,5555,9999,5555,5555,9999,5555,5555,9999,5555,5555
EFT$_\text{gs}$ I NNLO,23.8,0.4,0.4,4.7,0.4,0.4,9999,5555,5555,9999,5555,5555,9999,5555,5555,9999,5555,5555
EFT$_\star$ I NNLO,24.2,0.7,0.7,4.6,0.4,0.4,7.6,1.1,1,-2,16,15,23,19,60,9999,5555,5555
EFT$_\star$ II LO,36.4,0.4,0.4,9999,5555,5555,9999,5555,5555,9999,5555,5555,9999,5555,5555,2.45,0.02,0.02
EFT$_\star$ II NLO,22.4,0.5,0.5,6.1,0.5,0.5,9999,5555,5555,9999,5555,5555,9999,5555,5555,3.08,0.03,0.04
EFT$_\star$ II NNLO,26,0.5,0.5,4.6,0.4,0.4,7.7,1.2,1.1,7,29,44,-8,48,32,2.94,0.03,0.03
\end{filecontents*}
\csvreader[head to column names, late after line=\\]{paramsTable.csv}{}
{\ \Theory
& 
\ifthenelse{\equal{\ancA}{9999}}{---}{
\ifthenelse{\equal{\dancAup}{5555}}{$\ancA$}
{$\ancA^{+\dancAup}_{-\dancAdown}$}}
& 
\ifthenelse{\equal{\ancB}{9999}}{---}{
\ifthenelse{\equal{\dancBup}{5555}}{$\ancB$}
{$\ancB^{+\dancBup}_{-\dancBdown}$}}
& 
\ifthenelse{\equal{\ancC}{9999}}{---}{
\ifthenelse{\equal{\dancCup}{5555}}{$\ancC$}
{$\ancC^{+\dancCup}_{-\dancCdown}$}}
& 
\ifthenelse{\equal{\abb}{9999}}{---}{
\ifthenelse{\equal{\dabbup}{5555}}{$\abb$}
{$\abb^{+\dabbup}_{-\dabbdown}$}}
& 
\ifthenelse{\equal{\rab}{9999}}{---}{
\ifthenelse{\equal{\drabup}{5555}}{$\rab$}
{$\rab^{+\drabup}_{-\drabdown}$}}
& 
\ifthenelse{\equal{\betabb}{9999}}{---}{
\ifthenelse{\equal{\dbetaup}{5555}}{$\betabb$}
{$\betabb^{+\dbetaup}_{-\dbetadown}$}}
} 
\end{tabular}
\end{ruledtabular}
 \label{table:B8params}
\end{table*}  

\begin{figure}[tbh]
\begin{center}
\includegraphics[width=0.47\textwidth,clip=true]{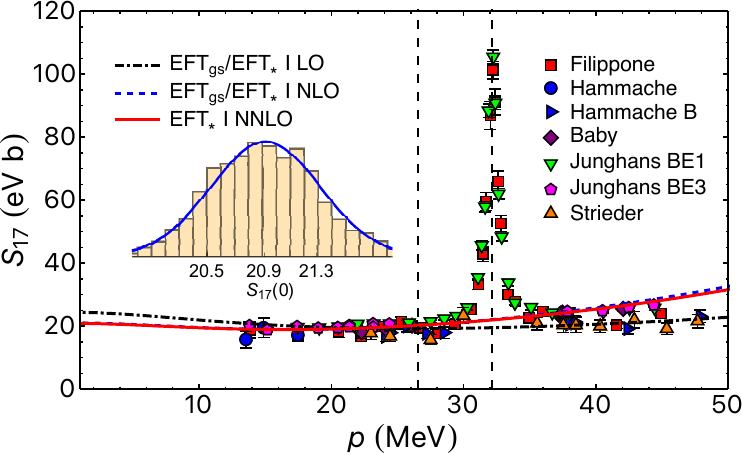}
\end{center}
\caption{\protect $S$-factor for \pBe ~in \EFTstarHead{}. Notation is the same as in Fig.~\ref{fig:E1CaptureEFTNNLO}. }
\label{fig:E1CaptureEFTstarNNLO}
\end{figure}

\begin{figure}[tbh]
\begin{center}
\includegraphics[width=0.47\textwidth,clip=true]{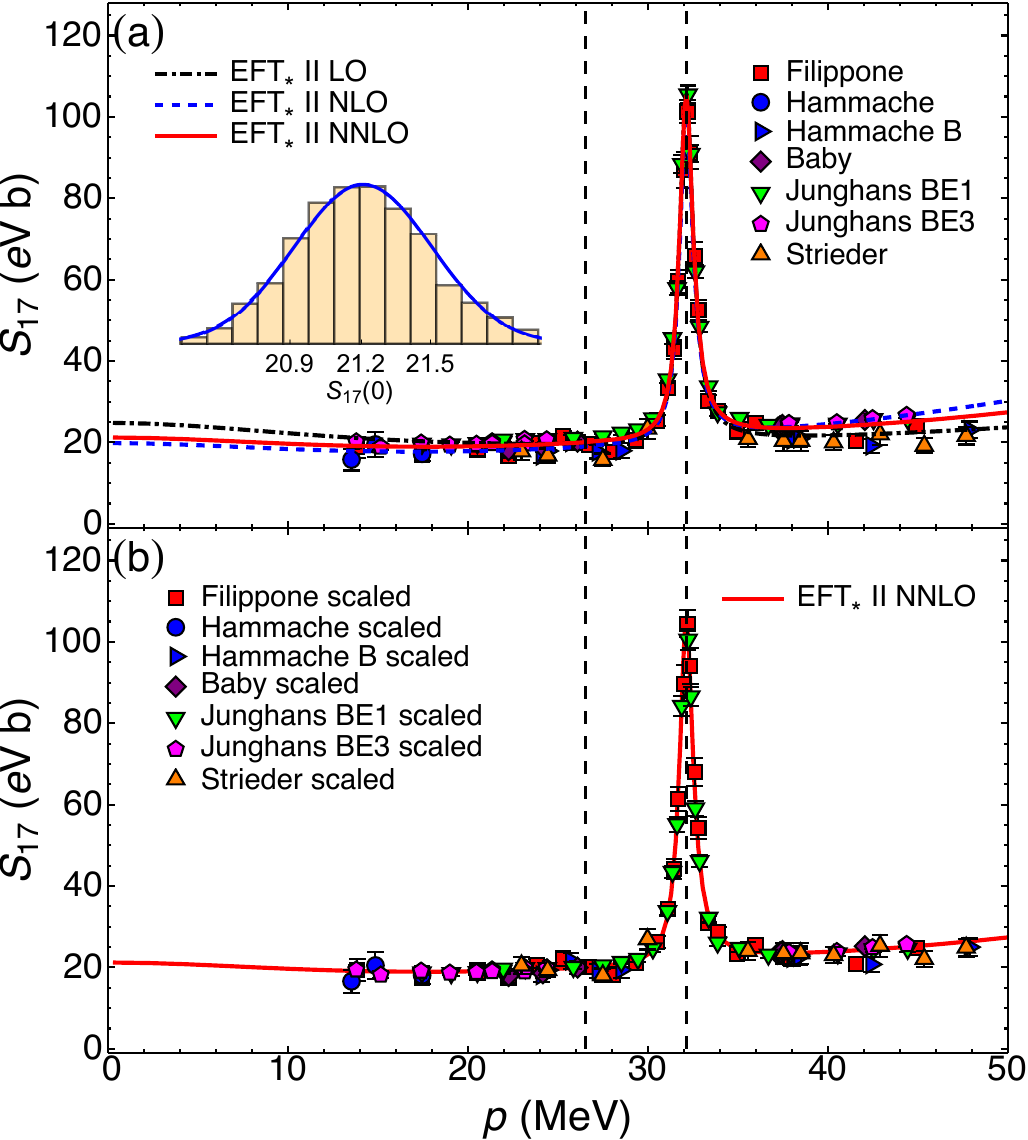}
\end{center}
\caption{\protect $S$-factor for \pBe ~in \EFTstarHead{} fitted in the momentum region II, $p\leq40.5$ MeV. Notation is analogous to the one in Fig.~\ref{fig:E1CaptureEFTNNLO}.  The bottom panel shows scaled data which is to be interpreted carefully as explained in the text. 
}
\label{fig:E1M1CaptureEFTstarNNLO}
\end{figure}

 The results from the fits to data in region I in \EFTgsHead{} are shown in Fig.~\ref{fig:E1CaptureEFTNNLO}. The curves show the mean values from the posterior distribution. The inset (also in Figs.~\ref{fig:E1CaptureEFTstarNNLO}, \ref{fig:E1M1CaptureEFTstarNNLO}) shows the 
 posterior distribution for the $S$-factor at threshold, and indicates the median and the interval containing 68\% of the posterior. The solid (blue) curve is a Gaussian fitted to the mean and s.d. of the distribution.
 We use Eq.~(\ref{eq:SfactorEFT}) without the M1 contribution and present the results up to NNLO. This calculation has a single fitting parameter $r_1^{(^5P_2)}$ (or $\mathcal Z^{(^5P_2)}$) at LO, and an additional parameter $\mathcal Z^{(^3P_2)}$ at NLO and NNLO, as shown in Table~\ref{table:B8params}.
We show the median and the interval that contains 68\% of the posterior distribution of the parameters. The fitted parameter has a size consistent with the power counting. 
 The $p$-wave effective momentum $r_1^{(^5P_2)}$ has a posterior that is normally distributed which is reflected in the symmetric error bands. 
It shows about 30\% variation in going from LO to NLO. We expect this order of variation even in the low momentum region I fit. The \EFTgsHead{} result has a non-zero NLO contribution at threshold from the $S=1$ spin channel. Thus when one goes from LO to NLO fits to the same data, the $r_1^{(^5P_2)}$ has to be adjusted for this NLO contribution. The NNLO contribution is small as discussed earlier and so the variation from NLO to NNLO parameters is small. The $\mathcal Z^{(^5P_2)}$ value is larger than the ANC estimate to match the capture data, as expected from Fig.~\ref{fig:CaptureEFTNNLO}.
 We find that the posteriors of the scaling factors $s_i$s are normally distributed and just show their mean values and s.d.s in Table~\ref{table:B8norms}.   Their sizes are as expected given the common-mode-error values. The data sets that have higher $S$-factor values are scaled down and the ones that have lower $S$-factor values are scaled up in the fits.     
 
 The results for \EFTstarHead{} in region I are shown in Fig.~\ref{fig:E1CaptureEFTstarNNLO}. 
 The curves show the mean values from the posterior distribution of the $S$-factor.
 We use Eq. (\ref{eq:corsssectionStar}), again without the M1 contribution. The momentum dependence in this theory is the same as in \EFTgsHead{} up to NLO.  At NNLO, initial state strong interactions in the $s$-wave necessitate fitting parameters $a_{22}$, $a_{12}$. We also fit  $\mathcal Z^{(^3P_2^\star)}$ due to the mixing with the $^3P_2^\star$ channel at this order in the expansion.  As discussed earlier, initial state $s$-wave strong interactions are small and as a result $a_{22}$, $a_{12}$ are not well constrained by capture data, Table~\ref{table:B8params}.   However, it also implies that these uncertainties have small effect on the $S$-factor at threshold. 
The scattering length $a_{22}$ is consistent with zero with large errors and the posterior for $a_{12}$ is bi-modal, however, the threshold  $S_{17}(0)$ calculated from marginalization over all the fit parameters is normally distributed as shown in the inset in Fig.~\ref{fig:E1CaptureEFTstarNNLO}.

 The \EFTgsHead{} I and \EFTstarHead{} I fits give very similar results. If we compare the evidences for the two fits  in region I at NNLO, we get~\cite{Nestle} $\ln[P(\mathrm{EFT}|D,H)/P(\mathrm{EFT_\star}|D,H) ]\approx \num{0.7+-0.3}$ that slightly favors \EFTgsHead{} I with 2 parameters over \EFTstarHead{} I with 5 parameters. However, the non-zero evidence is within the expected systematic error when comparing evidence calculations by different NS methods~\cite{Nestle,Brewer:2011,Premarathna:2019tup}.

\begin{table*}[tbh]
\centering
\caption{Mean and standard deviation of the posterior of the scaling factors $s_i$ for the 7 data sets: Filippone~\cite{Filippone:1984us}, Hammache~\cite{Hammache:2001}, Hammache~\cite{Hammache:1998}, Junghans BE1~\cite{Junghans:2010}, Junghans BE3~\cite{Junghans:2003bd,Junghans:2010} and Strieder~\cite{Strieder:2001}, respectively.}
\begin{ruledtabular}
\begin{tabular}{lrrrrrrr}
Theory 
& $s_1$ & $s_2$ & $s_3$ & $s_4$ & $s_5$ & $s_6$ & $s_7$\\ \hline
\begin{filecontents*}{normsTable.csv}
Theory,sa,dsa,sb,dsb,sc,dsc,sd,dsd,se,dse,sf,dsf,sg,dsg
EFT$_\text{gs}$/EFT$_\star$ I LO,1.03,0.03,1.09,0.06,1.04,0.03,1,0.02,0.94,0.01,0.98,0.01,1.14,0.03
EFT$_\text{gs}$/EFT$_\star$ I NLO,1.06,0.03,1.05,0.06,1.07,0.03,1.01,0.02,0.95,0.01,0.97,0.01,1.18,0.03
EFT$_\text{gs}$ I NNLO,1.06,0.03,1.04,0.06,1.07,0.03,1.01,0.02,0.95,0.01,0.97,0.01,1.18,0.03
EFT$_\star$ I NNLO,1.05,0.03,1.05,0.06,1.07,0.03,1.01,0.02,0.95,0.01,0.97,0.01,1.18,0.03
EFT$_\star$ II LO,1.01,0.02,1.1,0.06,1.05,0.03,0.94,0.01,0.94,0.01,0.96,0.01,1.1,0.02
EFT$_\star$ II NLO,0.99,0.02,1.01,0.06,1.06,0.03,0.99,0.01,0.92,0.01,0.93,0.01,1.13,0.02
EFT$_\star$ II NNLO,1.03,0.02,1.05,0.06,1.08,0.03,0.99,0.01,0.95,0.01,0.96,0.01,1.15,0.02
\end{filecontents*}
\csvreader[head to column names, late after line=\\]{normsTable.csv}{}
{\ \Theory
& 
\ifthenelse{\equal{\sa}{9999}}{---}{
\ifthenelse{\equal{\dsa}{5555}}{$\sa$}
{\SI{\sa+-\dsa}{}}}
& 
\ifthenelse{\equal{\sb}{9999}}{---}{
\ifthenelse{\equal{\dsb}{5555}}{$\sb$}
{\SI{\sb+-\dsb}{}}}
& 
\ifthenelse{\equal{\sc}{9999}}{---}{
\ifthenelse{\equal{\dsc}{5555}}{$\sc$}
{\SI{\sc+-\dsc}{}}}
& 
\ifthenelse{\equal{\sd}{9999}}{---}{
\ifthenelse{\equal{\dsd}{5555}}{$\sd$}
{\SI{\sd+-\dsd}{}}}
& 
\ifthenelse{\equal{\se}{9999}}{---}{
\ifthenelse{\equal{\dse}{5555}}{$\se$}
{\SI{\se+-\dse}{}}}
& 
\ifthenelse{\equal{\sf}{9999}}{---}{
\ifthenelse{\equal{\dsf}{5555}}{$\sf$}
{\SI{\sf+-\dsf}{}}}
& 
\ifthenelse{\equal{\sg}{9999}}{---}{
\ifthenelse{\equal{\dsg}{5555}}{$\sg$}
{\SI{\sg+-\dsg}{}}}
} 
\end{tabular}
\end{ruledtabular}
 \label{table:B8norms}
\end{table*}  

 Fig.~\ref{fig:E1M1CaptureEFTstarNNLO} shows our fits, mean values of the $S$-factor posterior, in region II of  \EFTstarHead{} expression in Eq.~(\ref{eq:corsssectionStar}) including the M1 contribution. We fit 
 $r_1^{(^5P_2)}$, $\mathcal Z^{(^3P_2)}$, 
 $\mathcal Z^{(^3P_2^\star)}$, $a_{22}$, 
 $a_{12}$ and  $\beta_{22}$ to capture data. The fitted values in Table~\ref{table:B8params} agree with the power counting estimates.
As in region I fits of \EFTstarHead{}, the posterior for $a_{12}$ is bi-modal. There is a correlation with $a_{22}$ whose posterior also shows a bi-modal distribution. Each modes are consistent with $|a_{12}|\sim|a_{22}|\sim 1/Q$ but the capture data is not sufficient to determine these parameters more precisely. 
The marginalized $S_{17}(0)$ is insensitive to this as seen in the posterior in the inset of  Fig.~\ref{fig:E1M1CaptureEFTstarNNLO}.
The curves are generated from the posterior of the fits instead of just the central values.  However, to give a rough estimate, in the bottom panel of the figure, we present an ``artist's" illustration of the fit where  we scaled the data by the values shown in Table~\ref{table:B8norms}, and added the errors from fits to the measurement errors in quadrature. It gives a sense of what the data sets would look like if they were scaled according to the respective scaling factors $s_i$s.

\begin{figure}[tbh]
\begin{center}
\includegraphics[width=0.47\textwidth,clip=true]{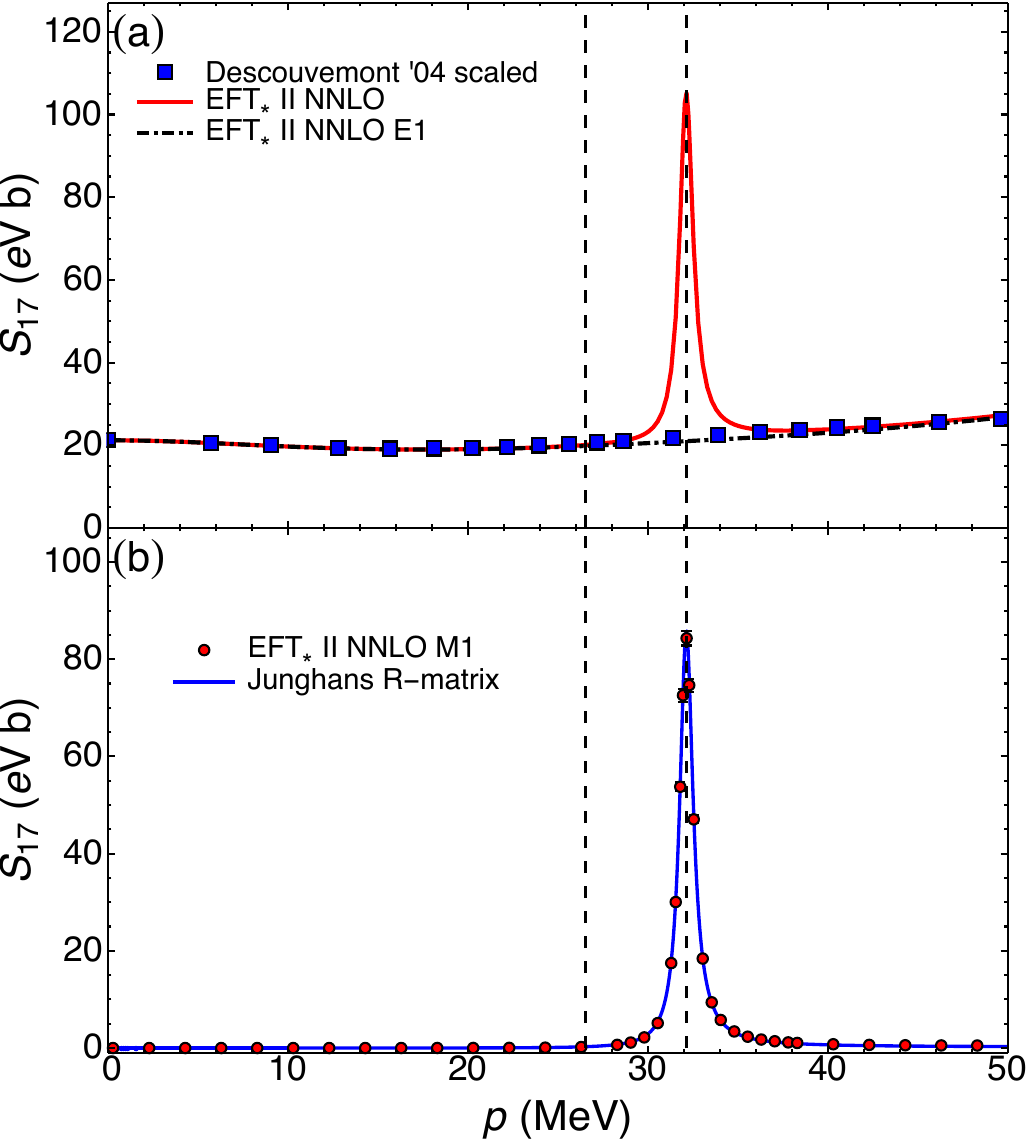}
\end{center}
\caption{\protect $S$-factor for \pBe ~from \EFTstarHead{} II fits at NNLO. The top panel compares the EFT result with numbers from Descouvemont~\cite{Descouvemont:2004} scaled by a factor of \descouvemont. The dot-dashed (black) curve is the EFT E1 contribution and the solid (red) curve is the total EFT E1+M1 contribution at NNLO. 
The bottom panel compares the EFT M1 numbers with the R-matrix calculation from Junghans {\em et al.}~\cite{Junghans:2003bd}.}
\label{fig:CompareEFT}
\end{figure}

 In Fig.~\ref{fig:CompareEFT}, we compare the \EFTstarHead{} II fit to the predictions from Descouvemont using the Minnesota nuclear interaction~\cite{Descouvemont:2004} that was used in the analysis in Solar II~\cite{Adelberger:2010qa}. We scaled Descouvemont's numbers by \descouvemont ~to match the EFT result at threshold. We also compare our resonance M1 contribution to a R-matrix calculation using the parameters used by Junghans {\em et al.}~\cite{Junghans:2003bd}. The R-matrix curve was used in one of the Solar II 
 analyses~\cite{Adelberger:2010qa}. The errors from the fits in the EFT results are displayed.  These comparisons show that the EFT results have reasonable agreement with known results.

\begin{figure}[tbh]
\begin{center}
\includegraphics[width=0.47\textwidth,clip=true]{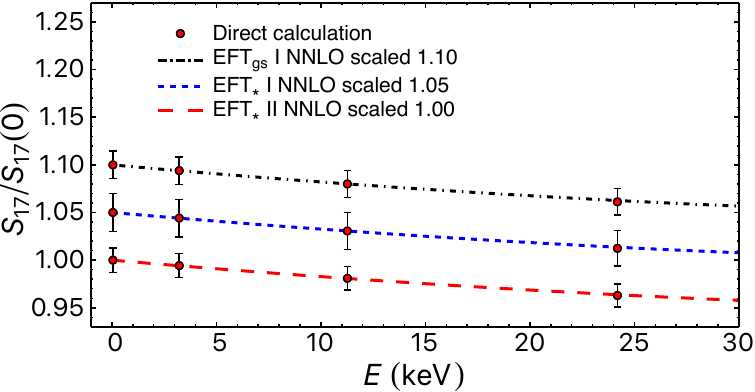}
\end{center}
\caption{\protect Taylor series approximation of $S_{17}(E)$ up to the quadratic term in $E$. 
The dot-dashed (black) curve is the \EFTgsHead{} I NNLO fit normalized to 1.1 times the threshold value, the dashed (blue) curve is the \EFTstarHead{}  I NNLO fit normalized to 1.05 times the threshold value, and the long-dashed (red) curve is the \EFTstarHead{} II NNLO fit normalized to the threshold value. The data-points (red circles) are the numerical evaluations scaled appropriately for the three fits.}
\label{fig:GamowPeak}
\end{figure}

\begin{table*}[tbh]
\centering
\caption{$S_{17}$ and its first two energy derivatives at $E_0=50\times10^{-3}$ keV. The first set of errors are from the fits. The second set is the estimated LO 30\%, NLO 10\% and NNLO 3\% EFT errors, respectively, from higher order corrections.}
\begin{ruledtabular}
\begin{tabular}{lrrr}
Theory & $S_{17}$ (\si{\eV\barn})
& $S'_{17}/S_{17}$ (\si{\mega\eV^{-1}})
& $S''_{17}/S_{17}$ (\si{\mega\eV^{-2}})\\ \hline
\begin{filecontents*}{S17TableA.csv}
Theory,S,dS,dSEFT,Sp,dSp,dSpEFT,Spp,dSpp,dSppEFT
EFT$_\text{gs}$/EFT$_\star$ I LO,24.4,3,73,-2.44,5,73,35.8,7,108
EFT$_\text{gs}$/EFT$_\star$ I NLO,21.1,3,21,-1.87,4,19,32.4,6,32
EFT$_\text{gs}$ I NNLO,20.7,3,6,-1.79,4,5,31.9,6,10
EFT$_\star$ I NNLO,20.9,4,6,-1.82,8,5,31.9,8,10
EFT$_\star$ II LO,24.8,3,74,-2.44,4,73,35.8,6,108
EFT$_\star$ II NLO,19.8,2,20,-1.91,3,19,32.7,5,33
EFT$_\star$ II NNLO,21.2,3,6,-1.89,4,6,31.9,6,10
\end{filecontents*}
\csvreader[head to column names, late after line=\\]{S17TableA.csv}{}
{\ \Theory & 
\num[parse-numbers=false,multi-part-units=brackets]{\S(\dS)(\dSEFT)}
&\SI[parse-numbers=false,multi-part-units=brackets]{\Sp(\dSp)(\dSpEFT)}{}
&\SI[parse-numbers=false,multi-part-units=brackets]{\Spp(\dSpp)(\dSppEFT)}{}
}
\end{tabular}
\end{ruledtabular}
 \label{table:B8S17}
\end{table*}  

The $S$-factor at threshold for the various EFT fits are shown in Table~\ref{table:B8S17}. We also list the first two energy derivatives calculated numerically. These values are as expected due to the peripheral nature of the capture reaction, Eq.~(\ref{eq:SfactorThreshhold}). The NLO result contains the $s$-wave capture without strong interaction in either channel and $d$-wave capture from $S=2$ channel. As a consequence the derivative ratio $S_{12}'(0)/S_{17}(0)$ is close to the one expected from Eq.~(\ref{eq:SfactorThreshhold}) expressed by Baye~\cite{Baye:2000}. The same holds for $S_{17}''(0)/S_{17}(0)$. NNLO brings in short range interactions in $s$-waves. However, as discussed earlier in Sec.~\ref{sec:PowerCounting}, NNLO and NLO results are expected to be similar. A direct calculation (using \EFTgsHead{} I NNLO parameters)  where we do not include any initial state strong interaction in EFT gives: $S_{17}'(0)/S_{17}(0)=\SI{-1.79+-0.03}{\mega\eV^{-1}}$, 
$S_{17}''(0)/S_{17}(0)=\SI{31.9+-0.6}{\mega\eV^{-1}}$ in close agreement with results with initial state strong interaction contributions in Table~\ref{table:B8S17}. Though we disagree with the treatment of the short range interaction in Ref.~\cite{Zhang:2015ajn} that goes beyond mere power counting arguments about how the perturbation should be organized---when they fit their form that has at least 5 more parameters than ours, say at NLO, to the same low energy data set it gives the same result as ours which is mostly Coulomb capture. It is not unreasonable since we fit to the same data. However, we speculate it is also indicative of the peripheral nature of the capture process. From our power counting perspective, short range interaction even if put incorrectly, are NNLO or higher effect. Thus when these short range interaction couplings are confronted with capture data in a parameter fit, these couplings take values such that the dominant contribution is the one from the $s$- and $d$-wave Coulomb capture without initial state strong interaction.

In Fig.~\ref{fig:GamowPeak},  we see a quadratic approximation using the first two energy derivatives gives an accurate description of the $S$-factor near the Gamow peak at $E\sim 20$ keV. The first derivative values in Table~\ref{table:B8S17} are comparable to the ones from Solar II~\cite{Adelberger:2010qa}. Our second derivatives are about a factor of 3 larger. However, we checked that reducing the second derivatives by a factor of 3 has very little effect on the $S_{17}$ value at the Gamow peak. We still get the curves well within the errors in the numerical evaluations indicated in Fig.~\ref{fig:GamowPeak}. For example, reducing 
$S_{17}''(0)/S_{17}\sim \SI{30}{\mega\eV^{-2}}$ to around $S_{17}''(0)/S_{17}\sim \SI{10}{\mega\eV^{-2}}$ changes the relative contribution of the quadratic term to $S$-factor at $E=\SI{30}{\kilo\eV}$ from $1.4\%$ to $0.5\%$, respectively.   We tabulate some $S$-factor values at various momenta (energies) in Table~\ref{table:S17long} of Appendix~\ref{sec:S17tabulation}.

We find the $S_{17}(0)$ posterior to be normally distributed in the fits, and therefore we just show the mean and s.d. of the distribution in Table~\ref{table:B8S17}. The theory error is estimated to be about 3\% in the NNLO E1 capture calculation and about 30\% in the LO M1 capture calculation. At threshold, M1 contribution is negligible so we estimate theory errors as 3\% from the E1 contribution. The error associated with the $s$-wave scattering lengths $a_0^{(1)}$, $a_0^{(2)}$ can also be included in this 3\% estimate using Eq.~(\ref{eq:SfactorThreshhold}) as a guideline. Within the context of Bayesian analysis, theory errors can be estimated~\cite{Furnstahl:2015rha,Melendez:2019izc}. This can be addressed in the  future. There does not seem to be a disagreement with the conservative EFT error estimates in the systems studied so far~\cite{Furnstahl:2015rha,Melendez:2019izc} to invalidate the power counting. We expect similar results for \pBe ~system.  The  recommended value from Solar II~\cite{Adelberger:2010qa} is $S_{17}(0)=\SI{20.8+-1.6}{\eV\barn}$. 
The three EFT evaluations of $S_{17}(0)$ at NNLO are consistent with this value and with each other within the errors. The current calculation has a smaller but comparable theory error than the earlier evaluation~\cite{Adelberger:2010qa}. The $S$-factor has been measured near the Gamow peak at Borexino~\cite{Tacaks:2018} $S_{17}(19^{+6}_{-5}\,\si{\kilo\eV})=\SI{19+-1.8}{\eV\barn}$ with an extrapolated value 
$S_{17}(0)=\SI{19.5+-1.9}{\eV\barn}$. The EFT calculations are in agreement with this result as well. Averaging the 3 fits, adding the fitting errors in quadrature and
then including the estimated NNLO theory error in quadrature to that gives \EFTSfactor.

\section{Conclusions}
 \label{sec:Conclusions}

We present a model-independent calculation of the \pBe ~using halo EFT. The $S$-factor for this reaction at threshold is estimated to a precision of 3\% from the combined experimental and theoretical error.  We present EFT calculations with and without the excited $^7\mathrm{Be}^\star$ core. The dominant E1 contribution is supplemented by the M1 capture through the $1^+$ resonance state of $^8$B for fitting the theory expressions to capture data over a wider energy region including the resonance. 

The E1 capture was calculated in a theory without (\EFTgsHead{}) and also with (\EFTstarHead{}) the excited $^7\mathrm{Be}^\star$ core as an explicit degree of freedom. 
In the latter theory, the  ${}^7{\rm Be}^{\star}$ core was explicitly 
accounted for in a coupled-channels 
formalism~\cite{Cohen:2004kf,Lensky:2011he,Higa:2020kfs}. Similar to the 
${}^7{\rm Li}(n,\gamma)^8{\rm Li}$ case~\cite{Higa:2020kfs}, we show that 
both theories have the same momentum dependence up to NLO. 
They differ in the interpretation of the final bound state in terms of short distance physics, c.f. Eqs.(\ref{eq:ERE-pwave})--(\ref{eq:Z-ere-relations}).
At NNLO, the two theories differ in their momentum dependence due to initial state strong interactions. 
Our formal expressions for E1 capture disagree with the ones from previous halo 
EFT studies~\cite{Zhang:2014zsa,Zhang:2015ajn,Zhang:2017yqc,Ryberg:2014exa} 
in similar ways as shown for the ${}^7{\rm Li}(n,\gamma)^8{\rm Li}$ 
case~\cite{Higa:2020kfs}. 
The M1 contribution is negligible except near the resonant energy of the $1^+$ excited state of $^8$B. We include the LO M1 capture which is identical in both the theories, and has negligible effect on the $S$-factor at threshold.

Based on the analysis of Sec.~\ref{sec:PowerCounting}, 
with particular attention to the relatively small contribution of the channel with spin $S=1$ compared to $S=2$, 
we establish the 
following power counting. 
The LO contribution consists of direct ({\em i.e.}, without initial state 
strong interactions, see for example,  Fig.~\ref{fig:DiagramA}) E1 capture from the 
initial ${}^5S_2$ state. 
NLO corrections come from direct E1 captures from the initial $d$-wave 
in the spin $S=2$ channel and initial $s$-wave in the spin $S=1$ channel. 
Initial state strong interactions enter only at NNLO in both channels---the 
loop contribution from Fig.~\ref{fig:DiagramB} due to rescattering scales as 
$a_0\,Q^2/\Lambda$ with Coulomb 
interactions~\cite{Higa:2016igc,Higa:2008dn,Luna:2019ufu,Schmickler:2019ewl}, 
in contrast with the non-Coulomb case~\cite{Higa:2020kfs}. In the $S=1$ 
spin channel, the unaturally large $a_0\sim 1/Q$ is compensated by a 
$Q/\Lambda$ suppression of this channel relative to $S=2$, which has a 
natural $a_0\sim 1/\Lambda$. 
Another NNLO correction comes from direct E1 capture from initial $d$-wave 
in the spin $S=1$ channel. 
That makes evident the peripheral nature of this capture reaction. 
Two-body current contribution in E1 capture is estimated to scale as $k_0 a_0 L_\mathrm{E1}$~\cite{Higa:2016igc,Premarathna:2019tup} with photon momentum $k_0=(p^2+\gamma^2)/(2\mu)\sim Q^3/\Lambda^2$. In the $S=2$ channel the $s$-wave scattering length scales as $a_0=a_0^{(2)}\sim 1/\Lambda$. Thus two-body current contributes at 
next-to-next-to-next-to-leading order (N$^3$LO) as $Q^3/\Lambda^3$. In the $S=1$ channel, the scattering length is larger $a_0=a_0^{(1)}\sim 1/Q$ so the two-body current contributes as $Q^2/\Lambda^2$. However, this spin channel starts one order higher in perturbation thus making the two-body current contribution also a N$^3$LO effect. 
The effective-range corrections to $s$-wave initial state strong interactions are also a N$^3$LO effect. 
M1 capture is included at LO, and mostly contributes around the $1^+$ resonance energy 
$\sim 600\,{\rm keV}$. Thus the estimated theory error at threshold is dominated by the $Q^3/\Lambda^3\sim 0.03$ correction
at N$^3$LO from the E1 transition.

The present calculation has only two free parameters up to NLO when fitted to capture data below 500 keV. At NNLO, \EFTgsHead{} still has two fit parameters and \EFTstarHead{} has 5.
In contrast, previous EFT 
calculations~\cite{Zhang:2015ajn,Zhang:2017yqc} need seven 
(nine with 
unconstrained $s$-wave scattering lengths) at NLO for 
capture below 500 keV. Their power 
counting were set to include short-distance two-body currents and effective 
ranges for E1 transition, a scenario slightly 
difficult to conciliate 
with the extremely peripheral nature of this reaction. We \emph{emphasize} that our disagreement with previous EFT calculations is not about just the power counting but also about the construction of the strong interaction operators themselves. We include an extra two-body M1 current when analysing capture data above 500 keV to describe the $1^+$ resonance contribution.  

We constrain our halo EFT parameters at each order with ``modern"  
direct capture 
data~\cite{Filippone:1984us,Hammache:2001,Hammache:1998,Baby:2003,Junghans:2010,Strieder:2001} 
using Bayesian analysis. That allows us to take into account 
common-mode-errors due to normalization of different data sets in a 
straightforward way. The fits were separated into two different energy ranges. 
Region I comprises data with energies $E\leq 500\,{\rm keV}$, thus, below the 
$1^+$ resonance energy $E_R=630\,{\rm keV}$. Both versions of halo EFT, \EFTgsHead{} and \EFTstarHead{}, were 
fitted in this region. There is no strong \emph{evidence} in the capture data to prefer one theory over the other in this energy region. 
In other words, both EFT$_{\star}$ and EFT$_\mathrm{gs}$ describe data equally well at $E\lesssim E_\star$ though EFT$_\mathrm{gs}$ does it with 3 fewer parameters. 
Region II includes data with $E\leq 1\,{\rm MeV}$. In region II, which includes energies above the threshold for the excited $^7\mathrm{Be}^\star$ core, only 
EFT$_{\star}$ is fitted, with the addition of the M1 capture, 
Eq.~(\ref{eq:sigmaM1}). 
Figs.~\ref{fig:E1CaptureEFTNNLO}-\ref{fig:E1M1CaptureEFTstarNNLO} present 
our results for $S_{17}$ as function of the c.m. momentum, 
confronted with 
data. 
The fitted EFT parameters summarized in Table~\ref{table:B8params} 
are consistent with the ones fixed from the ANCs. More importantly, the parameter estimates are consistent with the power counting. 
Table~\ref{table:B8S17} 
presents our results for $S_{17}(0)$, 
with the corresponding NNLO posterior distributions given by the insets 
in Figs.~\ref{fig:E1CaptureEFTNNLO}-\ref{fig:E1M1CaptureEFTstarNNLO}. $S_{17}$ for a range of momenta (energies) are included in Table~\ref{table:S17long} of Appendix~\ref{sec:S17tabulation}.

The non-resonant EFT results agree with the evaluation by Descouvemont~\cite{Descouvemont:2004} once it is scaled to match our  $S$-factor at threshold. The resonant contribution in the EFT also agrees with the R-matrix evaluation by Junghans {\em et al.}~\cite{Junghans:2003bd}. We provide the first two energy derivatives of the $S$-factor for a low-energy Taylor series extrapolation to the Gamow peak energy. 
The  NNLO EFT result for $S_{17}(0)$ averaged over the three fits (\EFTgsHead{}~I, 
\EFTstarHead{}~I, and \EFTstarHead{}~II) gives \EFTSfactor, 
in agreement with the Solar II 
recommended value of $20.8(1.6)\,{\rm eV\;b}$~\cite{Adelberger:2010qa} but with a much reduced theory error.

\begin{acknowledgments}
We thank B. Davids for discussing their work on analysis of capture data that was used in constructing the priors of the normalizations for the data sets used in the Bayesian analysis. We benefited from discussions with T. Papenbrock concerning scattering in shallow Coulomb systems. 
We also acknowledge useful discussions with K. Nollett, D. R. Phillips and X. Zhang. 
This work was supported in part by 
U.S. NSF grants PHY-1615092 and PHY-1913620 (PP, GR) 
and Brazilian agency FAPESP thematic projects No. 2017/05660-0 and No. 2019/07767-1, and INCT-FNA Proc. No. 464898/2014-5 (RH).
The $S$-factor figures for this article have been created using SciDraw~\cite{SciDraw}. 
\end{acknowledgments}

 \appendix
\section{Projectors}
\label{sec:Projectors}

The following are from Ref.~\cite{Fernando:2011ts,Higa:2020kfs} that we include for reference. 
For each partial wave we construct the corresponding projection operators 
from the relative core-nucleon velocity, the spin-1/2 
Pauli matrices $\sigma_{i}$'s, and the following spin-1/2 to spin-3/2 
transition matrices 
\begin{align}
S_1&=\frac{1}{\sqrt{6}}\begin{bmatrix}
-\sqrt{3} & 0 & 1 & 0\\
0&-1&0&\sqrt{3}
\end{bmatrix}\,, 
\nonumber\\
S_2 &= -\frac{i}{\sqrt{6}} \begin{bmatrix}
\sqrt{3} & 0 & 1 & 0\\
0&1&0&\sqrt{3}
\end{bmatrix}\,, 
\nonumber\\
S_3 &= \frac{2}{\sqrt{6}} \begin{bmatrix}
0 & 1 & 0 & 0\\
0&0&1&0
\end{bmatrix}\,,
\end{align}
which satisfy
\begin{align}
S_{i}S^{\dagger}_{j}&=\frac{2}{3}\delta_{ij}-\frac{i}{3}\epsilon_{ijk}
\sigma_{k}\,,\nonumber\\
S_{i}^{\dagger}S_{j}&=\frac{3}{4}\delta_{ij}-\frac{1}{6}\big\{J_{i}^{(3/2)},
J_{j}^{(3/2)}\big\}+\frac{i}{3}\epsilon_{ijk}J_{k}^{(3/2)}\,,
\end{align}
where $J_{i}^{(3/2)}$'s are the generators of the spin-3/2. 
We construct the Clebsch-Gordan coefficient matrices 
\begin{align}
F_i &=-\frac{i\sqrt{3}}{2}\sigma_2 S_i\, , \nonumber\\
Q_{i j} &= -\frac{i}{\sqrt{8}}\sigma_2\big(\sigma_i S_i+\sigma_j S_i\big),
\label{eq:ncorespinmatrices}
\end{align}
for projections onto spin channels $S=1$ and $S=2$, respectively. Then in 
coordinate space the relevant projectors that appear in the Lagrangians 
involving the $^7$Be ground state in Eqs. (\ref{eq:EFT}), (\ref{eq:EFTstar}) 
are~\cite{Rupak:2011nk,Fernando:2011ts}
\begin{align}
P_i^{(^3S_1)} &= F_j\, , 
\nonumber\\
P_{ij}^{(^5S_2)} &= Q_{ij}\,,
\nonumber\\
P^{(^3P_1)}_i &=\sqrt{\frac{3}{2} }F_x\P_y \epsilon_{i x y}\,, 
\nonumber\\
P^{(^3P_2)}_{i j} &=\sqrt{3} F_x \P_y\ R_{x y i j }\,,
\nonumber\\
P^{(^5P_1)}_i &=\sqrt{\frac{9}{5}} Q_{i x}\P_ x\,, 
\nonumber\\
P^{(^5P_2)}_{i j}&=\frac{1}{\sqrt{2}} Q_{x y} \P_z\ T_{x y z i j}\, . 
\label{eq:projdefr-space}
\end{align}
The tensors 
\begin{align}
R_{ijxy}&=\frac{1}{2}\left(\delta_{i x}\delta_{j y}+\delta_{i y}\delta_{j x} 
-\frac{2}{3}\delta_{i j}\delta_{x y}\right), 
\nonumber\\
T_{xyz i j}&=\frac{1}{2}\Big(\epsilon_{x z i}\delta_{y j}
+\epsilon_{x z j}\delta_{y i}
+\epsilon_{y z i}\delta_{x j}+\epsilon_{y z j}\delta_{x i} \Big),
\label{eq:RTGtensorsdef}
\end{align}
ensures total angular momentum $j=2$ is picked. 

The new projectors to describe the interactions in Eq.~(\ref{eq:EFTstar}) 
with the excited $^7\mathrm{Be}^\star$ core are 
\begin{align}
     P_i^{(^3S_1^\star)} & = -\frac{i}{\sqrt{2}}\sigma_2\sigma_i\, , \nonumber\\
    P_i^{(^3P_1^\star)} & = -i \frac{\sqrt{3}}{2}\sigma_2\sigma_x \P_y \epsilon_{i x y}\, , 
    \nonumber\\
    P_{ij}^{(^3P_2^\star)} &= -i\sqrt{\frac{3}{2}}\sigma_2\sigma_x \P_y R_{xyij}\, .
\end{align}

For the external states we introduce the photon vector 
($\varepsilon^{(\gamma)}_{i}$) 
and ground state $^8$B $2^+$ 
spin-2 ($\varepsilon_{ij}$) polarizations, obeying the 
following polarization  sums~\cite{Choi:1992,Fleming:1999ee}, 
\begin{align}
\sum_{\rm pol.}\varepsilon^{(\gamma)}_{i}\varepsilon^{(\gamma)*}_{j}&=
\delta_{ij}-\frac{k_ik_j}{k^2}\,, \nonumber\\
\sum_{\rm pol.\ ave.}\varepsilon_{ij}\varepsilon^{*}_{lm}
&=\frac{R_{ijlm}}{5}\,.
\label{eq:poltensordefs}
\end{align}

\section{Numerical \texorpdfstring{$S$-factor}{S-factor} values}
\label{sec:S17tabulation}
\begin{table*}[tbh]
\centering
\caption{$S$-factor as a function of c.m. momentum/energy from the NNLO Bayesian fits. Only the mean and the standard deviation of the posterior distribution of $S_{17}$ are shown.  Theory errors are not included.
}
\begin{ruledtabular}
\begin{tabular}{rrrrrr}
$p$ (\si{\mega\eV}) & $E$ (\si{\kilo\eV}) & \EFTgsHead{} I E1 $S_{17}$ (\si{\eV\barn})  & \EFTstarHead{} I E1 $S_{17}$ (\si{\eV\barn})& \EFTstarHead{} II E1 $S_{17}$ (\si{\eV\barn})
& \EFTstarHead{} II M1 $S_{17}$ (\si{\eV\barn})\\ \hline
\begin{filecontents*}{LongS17Table.csv}
momentum,energy,EFTI,dEFTI,EFTstarI,dEFTstarI,EFTstarIIE,dEFTstarIIE,EFTstarIIM,dEFTstarIIM
0.2864,0.05,20.72,0.27,20.92,0.39,21.21,0.27,0.0001804,0.0000034
2.3,3.224,20.6,0.27,20.81,0.39,21.09,0.27,0.0002031,0.0000038
4.3,11.27,20.34,0.27,20.54,0.38,20.81,0.27,0.0002617,0.0000049
6.3,24.19,19.99,0.26,20.17,0.37,20.43,0.26,0.0003737,0.0000069
8.3,41.98,19.61,0.26,19.79,0.35,20.01,0.25,0.0005763,0.00001
10.3,64.65,19.28,0.25,19.43,0.34,19.63,0.25,0.000941,0.000017
12.3,92.19,19.01,0.25,19.16,0.32,19.31,0.24,0.001604,0.000029
14.3,124.6,18.86,0.25,18.98,0.3,19.09,0.24,0.002826,0.000051
16.3,161.9,18.81,0.24,18.91,0.28,18.97,0.23,0.005131,0.000092
18.3,204.1,18.88,0.25,18.95,0.27,18.95,0.23,0.009605,0.00017
20.3,251.1,19.06,0.25,19.11,0.25,19.03,0.23,0.01865,0.00033
22.3,303,19.36,0.25,19.36,0.25,19.2,0.23,0.03807,0.00068
24.3,359.8,19.76,0.26,19.72,0.25,19.45,0.23,0.08376,0.0015
26.3,421.5,20.26,0.27,20.18,0.28,19.77,0.23,0.2086,0.0037
28.3,488.1,20.87,0.28,20.72,0.33,20.16,0.24,0.6578,0.012
29.05,514.3,9999,5555,9999,5555,20.31,0.25,1.134,0.02
29.8,541.2,9999,5555,9999,5555,20.47,0.26,2.191,0.039
30.55,568.7,9999,5555,9999,5555,20.63,0.31,5.134,0.091
31.3,597,9999,5555,9999,5555,20.8,0.57,17.48,0.31
31.55,606.6,9999,5555,9999,5555,20.85,0.87,30.04,0.53
31.8,616.2,9999,5555,9999,5555,20.91,1.5,53.75,0.95
31.95,622.2,9999,5555,9999,5555,20.95,1.9,72.56,1.3
32.15,630,9999,5555,9999,5555,20.99,2.2,84.3,1.5
32.3,635.8,9999,5555,9999,5555,21.02,2,74.65,1.3
32.55,645.7,9999,5555,9999,5555,21.08,1.3,47.04,0.83
33.05,665.6,9999,5555,9999,5555,21.19,0.6,18.41,0.33
33.55,685.9,9999,5555,9999,5555,21.31,0.4,9.402,0.17
34.05,706.5,9999,5555,9999,5555,21.43,0.33,5.777,0.1
35.3,759.4,9999,5555,9999,5555,21.73,0.29,2.62,0.046
36.55,814.1,9999,5555,9999,5555,22.04,0.28,1.617,0.028
37.8,870.7,9999,5555,9999,5555,22.38,0.29,1.167,0.02
40.3,989.7,9999,5555,9999,5555,23.1,0.31,0.7813,0.014
42.3,1090,9999,5555,9999,5555,23.75,0.34,0.6491,0.011
44.3,1196,9999,5555,9999,5555,24.46,0.37,0.5788,0.01
46.3,1306,9999,5555,9999,5555,25.24,0.4,0.5411,0.0093
48.3,1422,9999,5555,9999,5555,26.1,0.44,0.5231,0.0089
50.3,1542,9999,5555,9999,5555,27.03,0.49,0.5181,0.0088
\end{filecontents*}
\csvreader[head to column names, late after line=\\]{LongS17Table.csv}{}
{\ \momentum & \energy
& 
\ifthenelse{\equal{\EFTI}{9999}}{---}{
\ifthenelse{\equal{\dEFTI}{5555}}{$\EFTI$}
{\SI{\EFTI+-\dEFTI}{}}}
& 
\ifthenelse{\equal{\EFTstarI}{9999}}{---}{
\ifthenelse{\equal{\dEFTstarI}{5555}}{$\EFTstarI$}
{\SI{\EFTstarI+-\dEFTstarI}{}}}
& 
\ifthenelse{\equal{\EFTstarIIE}{9999}}{---}{
\ifthenelse{\equal{\dEFTstarIIE}{5555}}{$\EFTstarIIE$}
{\SI{\EFTstarIIE+-\dEFTstarIIE}{}}}
& 
\ifthenelse{\equal{\EFTstarIIM}{9999}}{---}{
\ifthenelse{\equal{\dEFTstarIIM}{5555}}{$\EFTstarIIM$}
{\SI[scientific-notation=false]{\EFTstarIIM+-\dEFTstarIIM}{}}}
} 
\end{tabular}
\end{ruledtabular}
 \label{table:S17long}
\end{table*}  

 We tabulate the numerical values of the $S$-factor at some momenta (and energies)  in Table~\ref{table:S17long}. Near the resonance momentum $p_R=32.15$ MeV where the M1 contribution is noticeable, we provide a finer momenta mesh points. For the EFT I NNLO and EFT$_\star$ I NNLO fits, the $S$-factor only contains contribution from E1 transition and they are applicable below $p_R$. For EFT$_\star$ II NNLO fit, we list the E1 and M1 contributions separately. We present the central values of different quantities to 4-significant figures. The c.m. momentum $p$ is converted to c.m. energy $E=p^2/(2\mu)$ with $\mu\approx820.49$ MeV.

\section{Coulomb Integrals}
\label{sec:Integrals}
The combination $B(p)+J_0(p)$ that appear in Fig.~\ref{fig:DiagramB} can be evaluated as
\begin{multline}
B(p)+J_0(p) =\frac{\mu}{3\pi}\frac{ip^3-\gamma^3}{p^2+\gamma^2}+k_C B'(p)
+\Delta B(p)\\
-\frac{k_C\mu}{2\pi}\left[2 H(\eta_p)+2\gamma_E -\frac{5}{3}+ \ln 4\pi
\right]\, .
\end{multline}
The function $B'(p)$ is given by the double integral
\begin{multline}
     B'(p)=\frac{\mu}{6\pi^2(p^2+\gamma^2)}\int_0^1dx\int_0^1dy
  \frac{1}{\sqrt{x(1-x)}\sqrt{1-y}}\\
  \times \left( xp^2\ln\left[\frac{\pi}{4k_C^2}
    (-yp^2+(1-y)\gamma^2/x-i0^+) \right] \right.\\
  \left. +p^2 \ln\left[\frac{\pi}{4k_C^2}
    (-yp^2-(1-y)p^2/x-i0^+) \right]  \right.\\
  \left. +x\gamma^2\ln\left[\frac{\pi}{4k_C^2}
    (y\gamma^2+(1-y)\gamma^2/x-i0^+) \right]  \right.\\
  \left. +\gamma^2\ln\left[\frac{\pi}{4k_C^2}(y\gamma^2-(1-y)
    p^2/x-i0^+) \right] 
  \right)\, ,
\end{multline}
which is reduced to a single integral before evaluating numerically.  The function $\Delta B(p)$ is obtained from 
\begin{multline}
  B_{ab}(p)  =-\frac{3}{\mu}\int d^3 r
  \left[\frac{G_C^{(1)}(-B;r',r)}{r'}\right]\Big|_{r'=0} \\
\times\frac{\partial G_C^{(+)}(E;\bm{r},0)}{\partial r_a}\frac{r_b}{r}
 \,  ,
 \end{multline}
 where
 \begin{multline}
  \left[\frac{G_C^{(1)}(-B;r',r)}{r'}\right]\Big|_{r'=0}
  =\left[\frac{G_C^{(1)}(-B;r,r')}{r'}\right]\Big|_{r'=0}\\
  =-\frac{\mu\gamma}{6\pi r}
  \Gamma\left(2+k_C/\gamma\right)
 W_{-\frac{k_C}{\gamma},\frac{3}{2}}(2\gamma r)\, ,
 \end{multline}
 and
 \begin{align}
 G_C^{(+)}(E;\bm{r},0) =-\frac{\mu}{2\pi r}\Gamma(1+i\eta_p)
  W_{-i\eta_p,\frac{1}{2}}(-i2 p r)\, .    
\end{align}
The integral $B_{ab}(p)$ is divergent at $r=0$. However, when combined with the contribution from $J(p)\delta_{ab}$ it is finite. Thus we make the substitution $(r_b/r)[\partial/\partial r_a]=(r_ar_b/r^2)[\partial/\partial r]
\rightarrow (\delta_{ab}/3)[\partial/\partial r]$ in the integral and accordingly 
$B_{ab}(p)\equiv B(p)\delta_{ab}$. The finite piece $\Delta B(p)$ is obtained numerically from $B(p)$ after subtracting the  zero and single photon contributions i.e. removing terms up to order $\alpha_e^2$~\cite{Higa:2016igc}.

The 1-loop magnetic moment contribution from Fig.~\ref{fig:M1capture} $(c1)$, $(c2)$ is given by the integral  
\begin{multline}
C_{ab}(p,\gamma)=
    \int\frac{d^3 m}{(2\pi)^3}\frac{d^3 l}{(2\pi)^3}\frac{d^3 q}{(2\pi)^3} m_a q_b \\
    \times G_c(-B,\bm m,\bm l) G_C(E,\bm l,\bm q)\\
    =9\int d^3 r\frac{r_a r_b}{r^2}\frac{G_C^{(1)}(-B,r',r)}{r'} 
    \frac{G_C^{(1)}(E,r,r'')}{r''}\Big|_{r'=0=r''}\, \\
    = 3\delta_{ab} \int d^3r   
    \frac{G_C^{(1)}(-B,r',r)}{r'} 
    \frac{G_C^{(1)}(E,r,r'')}{r''}\Big|_{r'=0=r''}\\
    =-\delta_{ab}\frac{i\mu^2\gamma p}{3\pi}\Gamma(2+i\eta_p)\Gamma(2+\frac{k_C}{\gamma})\\
 \times \int_0^\infty dr W_{-\frac{k_C}{\gamma},\frac{3}{2}}(2\gamma r) W_{-i\eta_p,\frac{3}{2}}(-i 2 p r)\\
    \equiv C(p,\gamma)\delta_{ab}\, .
\end{multline}
The function $C(p,\gamma)$ is evaluated as 
\begin{multline}
C(p,\gamma)=\mu^2\left[\frac{\lambda}{2\pi}+\frac{1}{3\pi}\frac{i p^3-\gamma^3}{p^2+\gamma^2}\right]\\
+\frac{k_C\mu^2}{3\pi}\left[
\frac{1}{D-4}+\frac{2}{3}+\gamma_E-\ln\frac{\pi\lambda^2}{k_C^2}
\right]\\
+k_C C'(p,\gamma)+\Delta C(p,\gamma)\, ,
\end{multline}
and
\begin{multline}
C'(p,\gamma)= \frac{\mu^2}{6\pi^2(p^2+\gamma^2)}\int_0^1 dx\int_0^1 dy \sqrt{\frac{x}{1-x}}\frac{1}{\sqrt{1-y}}\\
\times\left[p^2\ln\frac{-y p^2-(1-y)p^2/x-i 0^+}{4k_c^2}\right. \\
+p^2\ln\frac{-y p^2+(1-y)\gamma^2/x-i 0^+}{4k_c^2}\\
+\gamma^2\ln\frac{y \gamma^2-(1-y)p^2/x-i 0^+}{4k_c^2}\\
\left.+\gamma^2\ln\frac{y \gamma^2+(1-y)\gamma^2/x-i 0^+}{4k_c^2}
\right]\, .
\end{multline}
$\Delta C(p,\gamma)$ is evaluated from $C(p,\gamma)$ by subtracting the zero and single photon contributions similar to $\Delta B(p,\gamma)$. The divergences in $C(p,\gamma)$ are regulated by the two-body current coupling $L_{22}$. The regulated $\overline C(p,\gamma)$ is defined as 
\begin{align}
    \overline C(p,\gamma) = \frac{\mu^2}{3\pi}\frac{i p^3-\gamma^3}{p^2+\gamma^2}
    +k_C C'(p,\gamma)+\Delta C(p,\gamma)\, .
\end{align}

\setcounter{biburlnumpenalty}{9000}
\setcounter{biburllcpenalty}{7000}
\setcounter{biburlucpenalty}{8000}

\bibliographystyle{apsrev4-2}
%

\end{document}